\documentclass[traditabstract]{aa} 
\usepackage[a4paper, left=1.5cm, right=1.5cm, bottom=2.5cm, top=3.5cm]{geometry}
\usepackage{txfonts}
\usepackage{graphicx}
\usepackage{array}
\usepackage{natbib}
\usepackage{multirow}
\usepackage{relsize}
\usepackage{amsbsy}
\usepackage{hyperref}

\begin{document}

\title {The evolution of clustering length, large-scale bias and host halo mass at 2$<$z$<$5 in the VIMOS Ultra Deep Survey (VUDS)\thanks{Based on data obtained with the European Southern Observatory Very Large
Telescope, Paranal, Chile, under Large Program 185.A-0791.}}
\titlerunning{The evolution of clustering length, bias and halo mass at 2$<$z$<$5}

\author{A. Durkalec \inst{1}
\and O. Le F\`evre\inst{1}
\and A. Pollo\inst{19,20}
\and S. de la Torre\inst{1}
\and P. Cassata\inst{1,18}
\and B. Garilli\inst{3}
\and V. Le Brun\inst{1}
\and B.C. Lemaux \inst{1}
\and D. Maccagni\inst{3}
\and L. Pentericci\inst{4}
\and L.A.M. Tasca\inst{1}
\and R. Thomas\inst{1}
\and E. Vanzella\inst{2}
\and G. Zamorani \inst{2}
\and E. Zucca\inst{2}
\and R. Amor\'in\inst{4}
\and S. Bardelli\inst{2}
\and L.P. Cassar\`a\inst{3}
\and M. Castellano\inst{4}
\and A. Cimatti\inst{5}
\and O. Cucciati\inst{5,2}
\and A. Fontana\inst{4}
\and M. Giavalisco\inst{13}
\and A. Grazian\inst{4}
\and N. P. Hathi\inst{1}
\and O. Ilbert\inst{1}
\and S. Paltani\inst{9}
\and B. Ribeiro\inst{1}
\and D. Schaerer\inst{10,8}
\and M. Scodeggio\inst{3}
\and V. Sommariva\inst{5,4}
\and M. Talia\inst{5}
\and L. Tresse\inst{1}
\and D. Vergani\inst{6,2}
\and P. Capak\inst{12}
\and S. Charlot\inst{7}
\and T. Contini\inst{8}
\and J.G. Cuby\inst{1}
\and J. Dunlop\inst{16}
\and S. Fotopoulou\inst{9}
\and A. Koekemoer\inst{17}
\and C. L\'opez-Sanjuan\inst{11}
\and Y. Mellier\inst{7}
\and J. Pforr\inst{1}
\and M. Salvato\inst{14}
\and N. Scoville\inst{12}
\and Y. Taniguchi\inst{15}
\and P.W. Wang\inst{1}
}

\institute{Aix Marseille Universit\'e, CNRS, LAM (Laboratoire d'Astrophysique de Marseille) UMR 7326, 13388, Marseille, France
\and
INAF--Osservatorio Astronomico di Bologna, via Ranzani,1, I-40127, Bologna, Italy
\and
INAF--IASF, via Bassini 15, I-20133, Milano, Italy
\and
INAF--Osservatorio Astronomico di Roma, via di Frascati 33, I-00040, Monte Porzio Catone, Italy
\and
University of Bologna, Department of Physics and Astronomy (DIFA), V.le Berti Pichat, 6/2 - 40127, Bologna, Italy
\and
INAF--IASF Bologna, via Gobetti 101, I--40129, Bologna, Italy
\and
Institut d'Astrophysique de Paris, UMR7095 CNRS,
Universit\'e Pierre et Marie Curie, 98 bis Boulevard Arago, 75014
Paris, France
\and
Institut de Recherche en Astrophysique et Plan\'etologie - IRAP, CNRS, UniversitÃ© de Toulouse, UPS-OMP, 14, avenue E. Belin, F31400
Toulouse, France
\and
Department of Astronomy, University of Geneva,
ch. d'cogia 16, CH-1290 Versoix, Switzerland
\and
Geneva Observatory, University of Geneva, ch. des Maillettes 51, CH-1290 Versoix, Switzerland
\and
Centro de Estudios de F\'isica del Cosmos de Arag\'on, Teruel, Spain
\and
Department of Astronomy, California Institute of Technology, 1200 E. California Blvd., MC 249--17, Pasadena, CA 91125, USA
\and
Astronomy Department, University of Massachusetts, Amherst, MA 01003, USA
\and
Max-Planck-Institut f\"ur Extraterrestrische Physik, Postfach 1312, D-85741, Garching bei M\"unchen, Germany
\and
Research Center for Space and Cosmic Evolution, Ehime University, Bunkyo-cho 2-5, Matsuyama 790-8577, Japan
\and
SUPA, Institute for Astronomy, University of Edinburgh, Royal Observatory, Edinburgh, EH9 3HJ, United Kingdom
\and
Space Telescope Science Institute, 3700 San Martin Drive, Baltimore, MD 21218, USA 
\and
Instituto de Fisica y Astronomia, Facultad de Ciencias, Universidad de Valparaiso, Av. Gran Bretana 1111, Casilla 5030, Valparaiso, Chile
\and
Astronomical Observatory of the Jagiellonian University, Orla 171, 30-001 Cracow, Poland
\and
National Centre for Nuclear Research, ul. Hoza 69, 00-681, Warszawa, Poland
}


\abstract{
We investigate the evolution of galaxy clustering for galaxies in the redshift range $2.0<z<5.0$ using the VIMOS Ultra Deep Survey (VUDS). 
We present the projected (real-space) two-point correlation function $w_p(r_p)$ measured by using $3022$ galaxies with robust spectroscopic redshifts in two independent fields (COSMOS and VVDS-02h) covering in total $0.8$ $deg^2$. 
We quantify how the scale dependent clustering amplitude $r_0$ changes with redshift making use of mock samples to evaluate and correct the survey selection function. 
Using a power-law model $\xi(r) = (r/r_0)^{-\gamma}$ we find that the correlation function for the general population 
is best fit by a model with a clustering  length $r_0=3.95^{+0.48}_{-0.54} \textrm{ h}^{-1} \textrm{Mpc}$ and 
slope $\gamma=1.8^{+0.02}_{-0.06}$ at $z\sim2.5$, 
$r_0=4.35\pm0.60 \textrm{ h}^{-1} \textrm{Mpc}$  and $\gamma=1.6^{+0.12}_{-0.13}$ at $z\sim3.5$.
We use these clustering parameters to derive the large-scale linear galaxy bias $b_L^{PL}$, between galaxies and dark matter.
We find $b_L^{PL} = 2.68\pm0.22$ at redshift $z\sim3 $ (assuming $\sigma_8 = 0.8$), significantly higher than found
at intermediate and low redshifts.
We fit an HOD model to the data and we obtain that the average halo mass at redshift $z\sim3$ is $M_h=10^{11.75\pm0.23} \textrm{ h}^{-1} \textrm{M}_{\odot}$.
From this fit we confirm that the large-scale linear galaxy bias is relatively high at $b_L^{HOD} = 2.82\pm0.27$. 
Comparing these measurements with similar measurements at lower redshifts  
we infer that the star-forming population of galaxies at $z\sim3$ should evolve into the massive and bright ($M_r<-21.5$) galaxy population 
which typically occupy haloes of mass $\langle M_h\rangle = 10^{13.9} h^{-1} M_{\odot}$ at redshift $z=0$. 
}


\maketitle

\renewcommand{\arraystretch}{1.5}

\section{Introduction}

Since the very first galaxy surveys have revealed the complex structure of the Universe, the mapping of its evolution has become an important part of Cosmology.
Large surveys revealed that the Universe is composed of dense regions like clusters and filaments, but also of almost empty voids.
Additionally, the underlying dark matter structure and its evolution is shown to follow the visible baryonic matter, though the 
luminous matter is \textit{biased} in relation to the dark matter distribution.
It is still unclear how exactly this baryonic - dark matter relation looks like, especially in the early stages of galaxy formation, and how it changed through time.

The correlation function is a commonly used tool to describe how galaxies are clustered as a function of scale, and allows to put constrains on the evolution of galaxies (\citeauthor{Kauffmann1999} \citeyear{Kauffmann1999}, \citeauthor{Zehavi2011} \citeyear{Zehavi2011}).
It is based on the very simple idea of measuring the probability of finding two galaxies at a given redshift (\citeauthor{Peebles1980} \citeyear{Peebles1980}). 
The galaxy correlation function can be interpreted by introducing two kinds of approximations. 
The first and most extensively used formalism is based on a simple power-law approximation of the correlation function of the form $\xi(r)=(r/r_0)^{-\gamma}$ with two free parameters: the correlation length $r_0$, which describes how strongly galaxies are clustered, and the slope $\gamma$ (\citeauthor{DavisPeebles1983} \citeyear{DavisPeebles1983}).
The second, more recent and detailed approximation is based on halo occupation models (\citeauthor{Seljak2000} \citeyear{Seljak2000}, \citeauthor{Peacock2000} \citeyear{Peacock2000}, \citeauthor{Magliocchetti2003} \citeyear{Magliocchetti2003}, \citeauthor{Zehavi2004} \citeyear{Zehavi2004}, \citeauthor{Zheng2005} \citeyear{Zheng2005}). 
In this framework the correlation function is built from two components, which have their influence on different scales.
The \textit{One-halo} term dominates on small scales ($\lesssim 1.5 \textrm{ h}^{-1} \textrm{Mpc}$) and describes clustering of galaxies which reside within dark matter haloes. 
The \textit{two-halo} term  describes large scale ($\gtrsim 3 \textrm{ h}^{-1} \textrm{Mpc}$) galaxy clustering between different halos.  

Both formulations have been extensively used in the past years.
Correlation function measurements have been produced for most large galaxy surveys, such as 
the Sloan Digital Sky Survey (SDSS, \citeauthor{Connolly2002} \citeyear{Connolly2002}, \citeauthor{Zehavi2004} \citeyear{Zehavi2004}), 
2dF Galaxy Redshift Survey (2dFGRS, \citeauthor{Magliocchetti2003} \citeyear{Magliocchetti2003}), 
VIMOS-VLT Sky Survey (VVDS, \citeauthor{OLF2005} \citeyear{OLF2005}, \citeauthor{Pollo2006} \citeyear{Pollo2006}), 
VIMOS Public Extragalactic Survey (VIPERS, \citeauthor{Marulli2013} \citeyear{Marulli2013}), 
and DEEP2 (\citeauthor{Coil2006} \citeyear{Coil2006}).

Thanks to this remarkable effort it is now very well established that the strength of the galaxy clustering for the general population of galaxies is only mildly evolving from intermediate redshifts $z \sim 1$ to $z \sim 0$.
It has been found that galaxy clustering depends on a variety of galaxy properties like the luminosity, morphology, colour and spectral type of galaxies: luminous galaxies tend to be more clustered than faint ones and red galaxies with old stellar population are found to be more clustered than young blue ones 
(\citeauthor{Norberg2002} \citeyear{Norberg2002}, \citeauthor{Zehavi2004} \citeyear{Zehavi2004}, 2012, \citeauthor{Pollo2006} \citeyear{Pollo2006}, \citeauthor{delaTorre2007} \citeyear{delaTorre2007}, \citeauthor{Coil2008} \citeyear{Coil2008}, \citeauthor{Quadri2008} \citeyear{Quadri2008}, \citeauthor{Meneux2009} \citeyear{Meneux2006}, \citeyear{Meneux2008}, \citeyear{Meneux2009}, \citeauthor{Skibba2009} \citeyear{Skibba2009}, \citeauthor{Abbas2010} \citeyear{Abbas2010}, \citeauthor{Zehavi2011} \citeyear{Zehavi2011}, \citeauthor{Coupon2012} \citeyear{Coupon2012}).

At redshifts higher than $z\sim2$ the situation is less clear.
Various difficulties, mainly in collecting statistically significant and representative samples, need to be overcome.
While a number of attempts to measure galaxy clustering have been made at early epochs, interpreting results is not straightforward, mainly because of uncertainties connected to redshift determination, small volumes covered, and different galaxy populations selected from a range of methods.
Most of the measurements at high redshifts ($z>1$) are produced using photometric surveys targeting specific classes of galaxies or applying specific observation techniques, like B$z$K method (\citeauthor{Kong2006} \citeyear{Kong2006}, \citeauthor{Lin2012} \citeyear{Lin2012})  
or Lyman-break galaxy (LBG) selection (\citeauthor{Foucaud2003} \citeyear{Foucaud2003}, \citeauthor{Ouchi2004} \citeyear{Ouchi2004}, \citeauthor{Adelberger2005} \citeyear{Adelberger2005}, \citeauthor{Kashikawa2006} \citeyear{Kashikawa2006}, \citeauthor{Savoy2011} \citeyear{Savoy2011}, \citeauthor{Bielby2013} \citeyear{Bielby2013}).
In general, within a given sample and based on angular correlation function measurements, some evidence has been presented at $z > 1$ showing that clustering also seems to depend on luminosity, stellar mass or colour as observed on lower redshift, with bright galaxies clustering more strongly than faint ones (\citeauthor{Savoy2011} \citeyear{Savoy2011}) and passive galaxies clustering more strongly than star-forming galaxies at a given stellar mass (\citeauthor{Lin2012} \citeyear{Lin2012}).
However the relation of galaxy samples used in these analyses to the general population of galaxies is not well established,
which makes the study of evolution of the galaxy clustering at $z>2$ difficult.
Moreover these galaxies can not be easily connected with the galaxy populations at lower redshifts, which make it difficult to conduct a consistent study of galaxy clustering evolution from high redshift to $z=0$.

From the theoretical point of view, the correlation function should evolve with time - $\xi = \xi(r,t)$, because the density field of the Universe evolves over time.
In the framework of the Newtonian linear perturbation theory (and in the matter-dominated era of the history of the Universe) in the co-moving coordinates 
the density contrast can be decomposed into time-dependent and
space-dependent factors: $\delta(\vec{r},t) = D(t) \delta_0(\vec{r})$,
where $\delta_0(\vec{r})$ is the present day value of the density contrast
at a given location, and $D(t)$ is referred to as the growth factor, which depends on the parameters of the assumed cosmological models.
In practice, it means that the spatial shape of the density fluctuations
in co-moving coordinates does not change, and only their amplitude increases.
Consequently, the amplitude of the correlation function should increase over time: $\xi(r,t) = D^2(t) \xi(r,t_0)$, with a dependence on the cosmological parameters (\citeauthor{Peebles1980} \citeyear{Peebles1980}, \citeauthor{Schneider2006} \citeyear{Schneider2006}).

The consistent measurement of the evolution of a correlation
function amplitude for a given galaxy population is therefore enough - in principle -
to test the paradigm of the gravitational perturbations
as the origin of the large scale structure of the Universe (and a way to estimate cosmological parameters). 
Extending these measurements to as high redshifts as possible
is necessary both for the theoretical framework of the large scale structure
evolution and for the galaxy formation and evolution models.

\begin{figure*}[t!]
 \centering
 \begin{tabular}{cc}
 \includegraphics[width=0.45\textwidth, height=0.45\textwidth, angle=270]{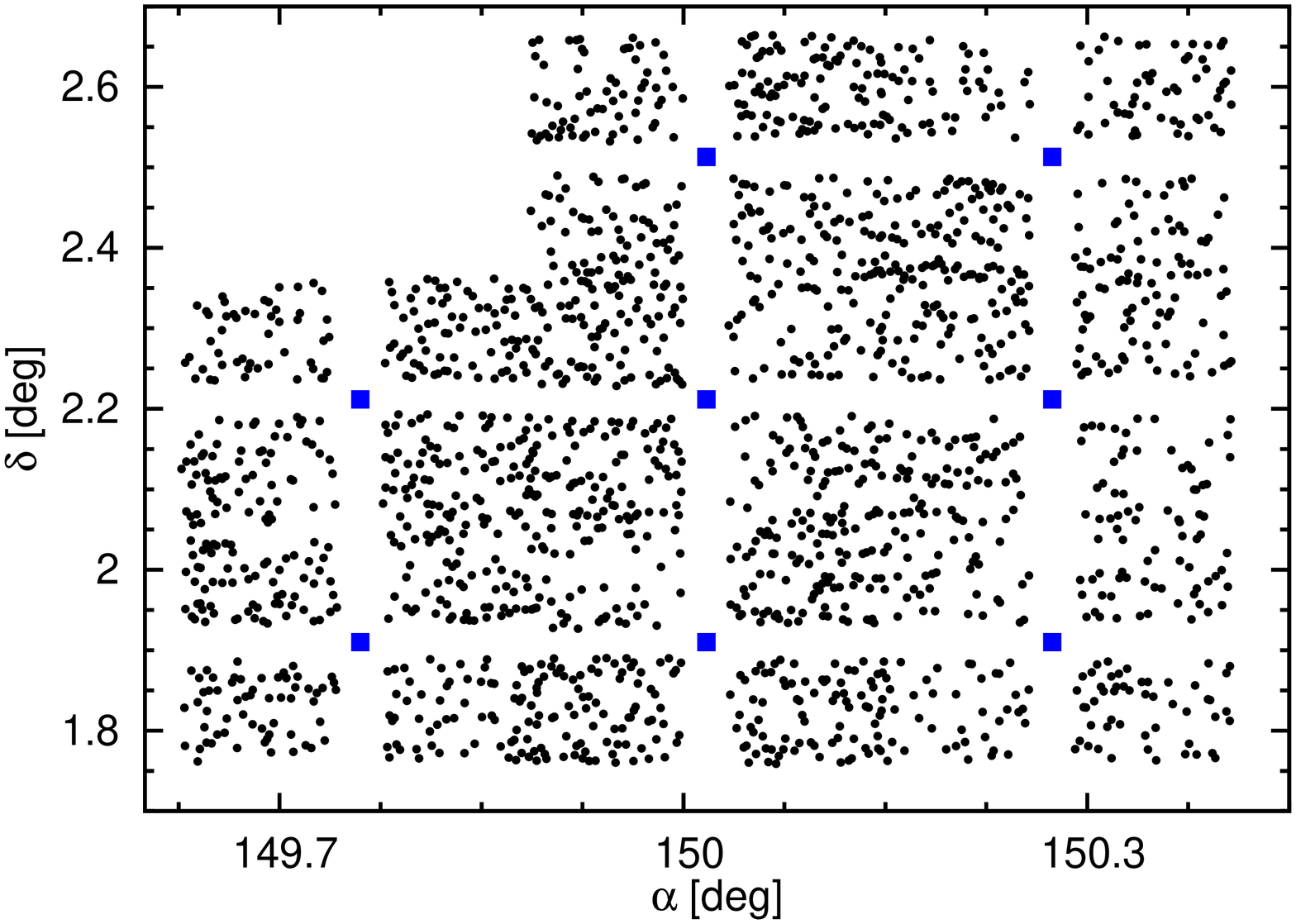} &  \includegraphics[width=0.45\textwidth, height=0.45\textwidth, angle=270]{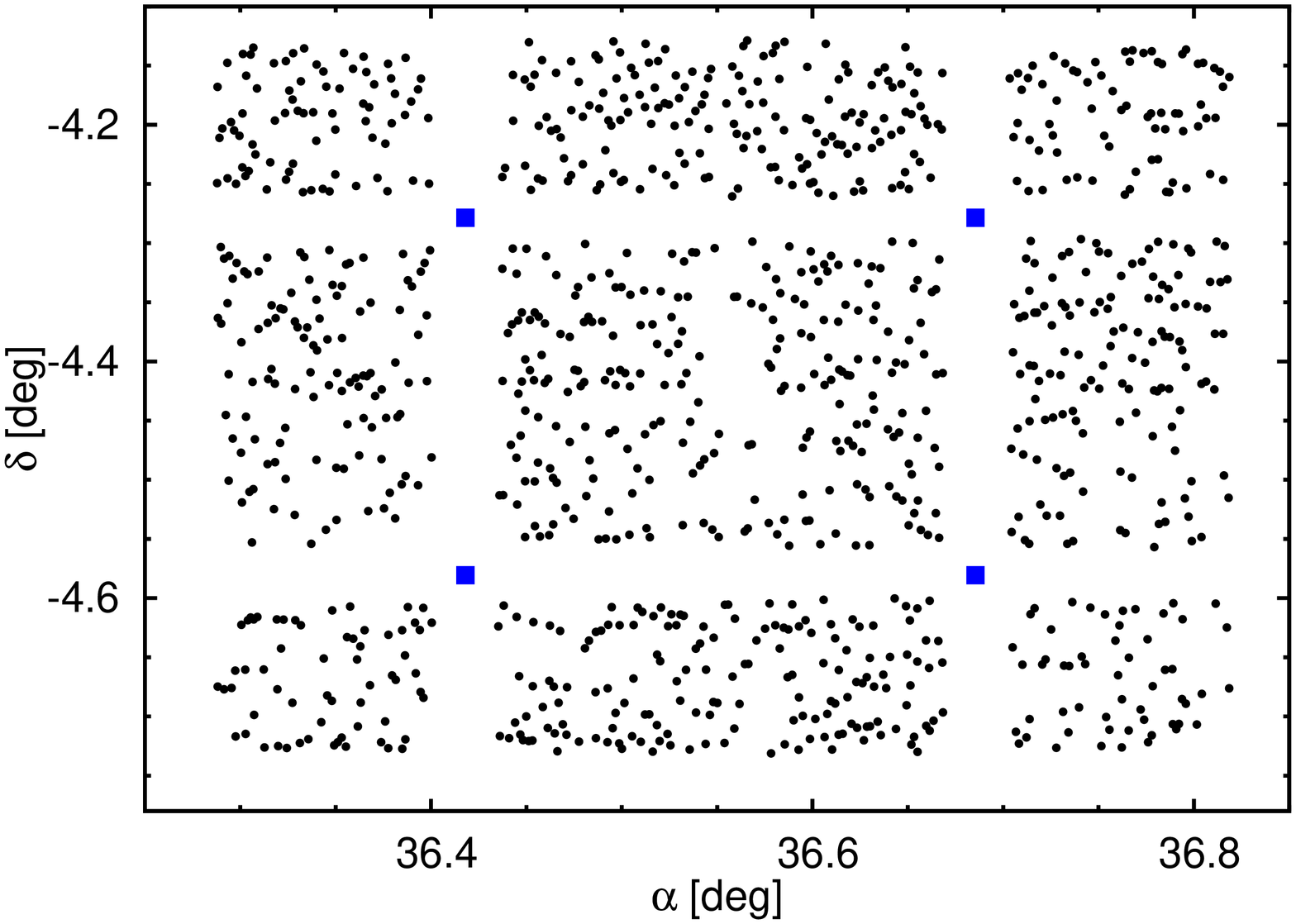} 
 \end{tabular} 
 \caption{The distribution of galaxies with spectroscopic redshifts $2<z<5$ in two independent VUDS fields: COSMOS on the \textit{left} panel and VVDS-02h on \textit{right} panel.
	  The blue squares indicates VIMOS pointing centers.}
 \label{fig:field_galaxy}
\end{figure*}

In this work we attempt to improve our current view of the evolution of the clustering of galaxies at $z>2$,
investigating how galaxy clustering evolved from the early phases of galaxy assembly to present times.
We present a clustering analysis of the VIMOS Ultra Deep Survey (VUDS), the largest spectroscopic survey covering the redshift range $2<z<6$ in a continuous way to date (Le F\`evre et al. 2014), using a sample of $\sim3000$ galaxies with  confirmed spectroscopic redshifts in the range $2<z<5$. 
The sample covers a total area of $0.81$ $deg^2$ observed in two independent fields, COSMOS and VVDS-02h, with a mean redshift value of $z\sim3$.
We use a halo occupation distribution (HOD) model to put constrains on the properties of dark matter halos hosting star-forming galaxies at redshifts $z\sim3$.

The paper is organized as follows. 
In section \ref{sec:2_data} we briefly describe the properties of the VUDS survey and our selected samples.
The method used to measure the correlation function and derive power-law and HOD fits is presented in section \ref{sec:3_methods}.
Results are described in section \ref{sec:4_results}. 
We discuss our findings and compare to other work, both at low and high redshifts, in section \ref{sec:5_discussion}, and we summarize our work in section \ref{sec:6_summary}.

We adopt a flat $\Lambda CDM$ cosmological model with $\Omega_m = 0.25$, $\Omega_{\Lambda}=0.75$ (we note that using the latest cosmology parameters from e.g. \citeauthor{Planck2013} \citeyear{Planck2013} does not change  conclusions drawn from clustering analysis, see \cite{Zehavi2011}). 
The Hubble constant is normally parametrized via $h = H_0/100$ to ease comparison with previous works, while a value $H_0=70$ km $\textrm{s}^{-1}$ Mpc is used when computing absolute magnitudes and stellar masses.
We report correlation length measurements in comoving coordinates and express magnitudes in the AB system.


\section{The Data}
\label{sec:2_data}

\subsection{The VUDS Survey summary}
\label{sec:vuds_summary}
The VIMOS Ultra Deep Survey (VUDS, \citeauthor{OLF2014} \citeyear{OLF2014}) is a spectroscopic survey targetting $\sim10,000$ galaxies performed with the VIMOS multi-object spectograph (\citeauthor{OLF2003} \citeyear{OLF2003}) at the European Southern Observatory Very Large Telescope.
The main aim  of the survey is to study early phases of galaxy formation and evolution at $2<z<6$.
The survey covers a total area of $1$ $deg^2$ in three independent fields, reducing the effect of cosmic variance, an important property for galaxy clustering measurements. 
Targets are selected based on photometric redshifts $(z_{phot}>2.4+1 \sigma)$ derived from deep multi-band photometry available in the VUDS fields, augmented by colour selection (e.g. Lyman Break Galaxies, LBG) when not selected by the photometric redshift criterion, and limit of $i_{AB}=25$ is imposed.
Details about the survey strategy, target selection, as well as data processing and redshift measurements are presented in ~\cite{OLF2014}. 
Here we briefly describe VUDS features which are relevant for our work.

  \begin{table}[t]
    \begin{center}
     \caption{The number of galaxies in VUDS fields for each reliability flag in the redshift range $2<z<5$, as used in this study.}
      \scalebox{1.2} {
	\begin{tabular}{c|c|c|c|c|c} \hline \hline 
	  \multirow{2}{*}{VUDS field}	&	\multicolumn{4}{c|}{reliability flag $zflag$}  & \multirow{2}{*}{Area (deg$^2$)} \\ \cline{2-5}
					&	2	&	3	&	4	&	9   &	     \\ \hline
	    COSMOS			&	939	&	607	&	322	&	72  &	0.50 \\
	    VVDS-02h			&	471	&	384	&	207	&	20  &	0.31 \\ \hline
	    Total			&	1410	&	991	&	529	&	92  &	0.81 \\ \hline
	\end{tabular}
     \label{table:sample}
      }
      \end{center}
   \end{table}

Redshift measurements were carried out in a way similar to that developed for the VVDS survey (\citeauthor{OLF2005} \citeyear{OLF2005}, \citeauthor{OLF2013} \citeyear{OLF2013}), the zCOSMOS survey (\citeauthor{Lilly2007} \citeyear{Lilly2007}), and VIPERS survey (\citeauthor{Guzzo2014} \citeyear{Guzzo2014}). 
The core engine for redshift measurement is cross-correlation of the observed spectrum with reference templates using the EZ redshift measurement code (\citeauthor{Garilli2010} \citeyear{Garilli2010}). 
At the end of the process, each redshift measurement is assigned a reliability flag which expresses the reliability of the measurement:
\begin{itemize}
 \item Flag 0: No redshift could be assigned
 \item Flag 1: $50-75\%$ probability to be correct
 \item Flag 2: $75-85\%$ probability to be correct
 \item Flag 3: $95-100\%$ probability to be correct
 \item Flag 4: $100\%$ probability to be correct
 \item Flag 9: $\sim80\%$ probability to be correct; spectrum has a single emission line
\end{itemize}

The VUDS sample benefits from an extended multi-wavelength dataset (see \citeauthor{OLF2014} \citeyear{OLF2014}). 
The multi-wavelength photometry is used to compute absolute magnitudes and stellar masses from SED fitting using the 'Le Phare' code (\citeauthor{Arnouts1999} \citeyear{Arnouts1999}, \citeauthor{Ilbert2006} \citeyear{Ilbert2006}), 
{as described in detail by \cite{Ilbert2013} and references therein.

\subsection{The sample selection for clustering analysis}

In this work we are using only the objects with assigned reliability flag $zflag=2,3,4,9$ in the redshift range $2<z<5$.
We use two independent fields, COSMOS and VVDS-02h, covering a total area $0.81$ $deg^2$, as data for the third VUDS field in the ECDFS are still too sparse to be used for clustering analysis. 
Covered area corresponds to a volume $\sim 3 \times 10^7$Mpc$^3$ sampled in the redshift range $2<z<5$.
Our sample consists of $3022$ galaxies with reliable spectroscopic redshifts in the range $2<z<5$.
The sample is summarized in Table \ref{table:sample}.

The spatial distribution of the spectroscopic galaxy sample in each of these fields is presented in Figure \ref{fig:field_galaxy}, while Figure \ref{fig:redshift_distribution} presents their redshift distributions.
Mean redshift value for the whole sample is $z\sim2.96$.
We use 3 different redshift bins to study our sample. 
In a first sample we include the whole star-forming galaxy population observed by VUDS in the range $2<z<5$ with a mean redshift $z\sim2.96$. We also define a \textit{low} redshift sample with a mean redshift $\bar{z}=2.5$ in the range z=[2,2.9] and a \textit{high} redshift sample with $\bar{z}=3.45$ in z=[2.9,5], which roughly represent two equal cosmic time bins.
The general properties of these samples including median luminosity $M_U^{median}$ and stellar mass $\log M_*^{median}$  are listed in Table \ref{tab:cf_redshift}.

Because of the $i_{AB}=25$ limit of the VUDS selection,
the galaxies in the high redshift sample are $\sim 0.5 \ \textrm{mag}$ brighter and $\sim0.2 \ M_{\odot}$ more massive than those in the low redshift sub-sample, which is the expected effect taking into account the VUDS selection strategy.
The impact of these differences on the clustering measurements of our three samples are fully discussed in Section \ref{sec:5_discussion}.

\begin{figure}
 \centering
 \includegraphics[height =0.5\textwidth, width =0.5\textwidth, angle=270]{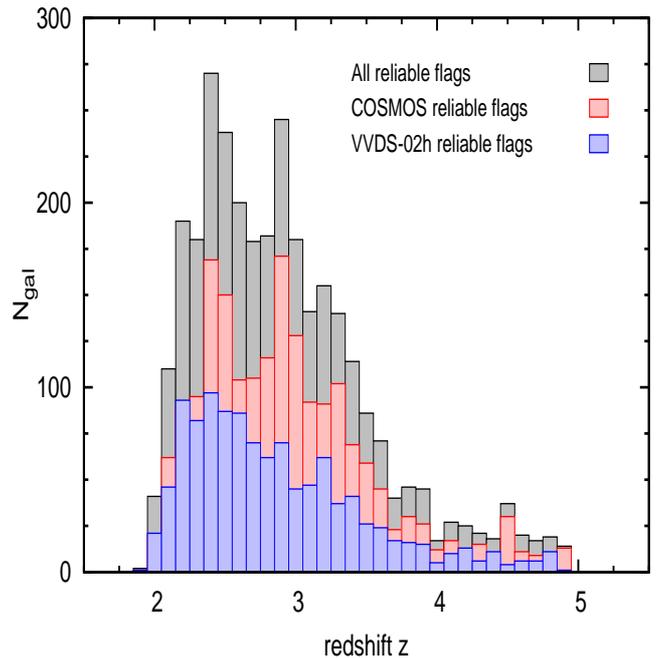}
 \caption{The redshift distribution of the used VUDS galaxy sample for redshift range $2.0<z<5.0$. 
	  The filled gray histogram represents total sample of reliable galaxies (with $zflag = 2-4,9$, $\sim 80\%$ reliability), while the red and blue histograms represent the contribution from COSMOS and VVDS2h fields respectively.}
 \label{fig:redshift_distribution}
\end{figure}

\subsection{VUDS mock catalogues}
\label{sec:mock}
To estimate uncertainties of the correlation function (see Sec. \ref{sec:errors}) and to test robustness (see App. \ref{sec:corr_correction}) of our clustering measurements we make use of a large number of mock galaxy samples, which are designed to mirror the VUDS sample in the range $2<z<5$ in the most realistic way.  

The 66 independent mock samples have been created based on a method which includes the halo occupation distribution (HOD) and the Stellar-to-Halo Mass Relation (SHMR).
This method is the same as developed and introduced by \cite{delaTorre2013}, and we refer the reader to the details presented in this paper.
We provide below a brief summary of its most important features.

In the first step the mock catalogues containing stellar masses have been created from the MultiDark $N$-body simulation (\citeauthor{Prada2012} \citeyear{Prada2012}) and Pinocchio halo lightcones (\citeauthor{Monaco2002} \citeyear{Monaco2002}).
We followed the stellar mass to halo mass ratio (SMHR) approach based on the assumption of a monotonic relation between halo/subhalo masses and the stellar masses of the galaxies associated with them.
We first populated the haloes in lightcones with subhaloes. 
For this we randomly distributed subhaloes around each distinct halo following a NFW profile (\citeauthor{NFW1997} \citeyear{NFW1997}), so that their number density satisfies the subhalo mass function proposed by \cite{Giocoli2010}.
Then we assigned a galaxy to each halo and subhalo, with a stellar mass given by the SHMR of \cite{Moster2013}.

The procedure which was followed in the next step is similar to the one used in the VVDS, zCOSMOS and VIPERS surveys also based on VIMOS observations as the VUDS survey 
(\citeauthor{Meneux2006} \citeyear{Meneux2006}, \citeauthor{Iovino2010} \citeyear{Iovino2010}, \citeauthor{delaTorre2011b} \citeyear{delaTorre2011b}, \citeyear{delaTorre2013}).
In order to obtain fully realistic VUDS mock catalogues we add the detailed survey selection criteria (see Sec. \ref{sec:vuds_summary} and \citeauthor{OLF2014} \citeyear{OLF2014}).
This procedure in the end produces mock parent photometric galaxy catalogues.
We then apply the slit-positioning algorithm (SSPOC) with the same settings as used for the VUDS survey (\citeauthor{Bottini2005} \citeyear{Bottini2005}).
This allowed us to reproduce the VUDS footprint on the sky for a proper simulation of the small-scale angular incompleteness in the mocks.

As a result of this procedure, 66 realistic mock galaxy catalogues have been produced. They contain the detailed survey completeness function and observational biases of the VUDS survey.


\section{Clustering measurements}
\label{sec:3_methods}

\subsection{Two-point correlation function}
The two point correlation function is one of the most frequently used statistical tool used to measure and investigate galaxy clustering. 
It measures the probability, above Poisson, of finding two galaxies separated by a given distance $r$ (\citeauthor{Peebles1980} \citeyear{Peebles1980}).
  \begin{equation}
   P = [1+\xi(r)]\rho \textrm{d}V_1\textrm{d}V_2,
  \end{equation}
where $P$ is the probability of finding these two galaxies in two infinitely small 
volumes $dV_1$ and $dV_2$ separated by the distance $r$, while the average density of galaxies is given by $\rho$.

Although the idea is relatively simple, this definition cannot be straightforwardly applied to compute the correlation function from real data samples, because of the limitations of galaxy surveys themselves. 
In principle, to retrieve ideal correlation functions one would need an unlimited survey which covers the whole sky, and includes all galaxies.
Naturally, in practice such surveys are unreachable.
A number of estimators of $\xi(r)$, aimed at minimizing the effects related to the limited number of objects and limited areas covered by available 
surveys, have been proposed (see, e.g., \citeauthor{DavisPeebles1983} \citeyear{DavisPeebles1983}, \citeauthor{Hamilton1993} \citeyear{Hamilton1993}).
The estimator the most commonly used because of its well established capability to minimize the above mentioned observational limitations is the \cite{LandySzalay1993} estimator, which we also apply in this work:
  \begin{equation}
  \label{eq:landy_szalay}
   \xi(r) = \frac{N_R(N_R-1)GG(r)}{N_G(N_G-1)RR(r)} - 2 \frac{(N_R-1)GR(r)}{N_G RR(r)}+1
  \end{equation}
Here, $N_G$ and $N_R$ represent, respectively, numbers of galaxies in the galaxy sample and randomly distributed objects in the same volume as observed in the survey; 
$GG(r)$ is the number of distinct galaxy-galaxy pairs with separations lying in the interval $(r,r+dr)$.
Similarly, $RR(r)$ and $GR(r)$ are the numbers of random-random pairs and galaxy-random pairs, respectively, in the same intervals. 
The galaxy pairs $GG$, $GR$ and $RR$ are normalized by $N_G(N_G-1)$, $N_GN_R$ and $N_R(N_R-1)$ respectively, where $N_G$ and $N_R$ are the number of galaxies in the data and random catalogues. 

In this work we make galaxy clustering measurements using combined data from two independent VUDS survey fields - COSMOS and VVDS2h. 
The final $\xi^{final}(r)$ is computed for all fields simultaneously by using Landy\& Szalay estimator (Eq. \ref{eq:landy_szalay}), and the differences in size and numbers between the fields are accounted for by an appropriate weighting scheme. 
In particular, each pair is multiplied by the number of galaxies per unit volume for the given field:
  \begin{equation}
  \label{eq:composite_cf}
   \xi^{final}(r) = \frac{\mathlarger{\sum\limits_{i=1}^{n_{field}}}w_i\cdot\Big(GG_i(r)-2GR_i(r)+RR_i(r)\Big)}{\mathlarger{\sum\limits_{i=1}^{n_{field}}}w_i\cdot RR_i(r)},
  \end{equation}
where $w_i = \left(N_{gal,i}/V_i \right)^2$. 

The Landy \& Szalay estimator requires the creation of a random catalogue, which follows the geometrical properties of the corresponding galaxy sample and is distributed in the same volume (or on the same area on the sky, in its angular version). 
The number of objects generated within this catalogue is also crucial - in practice it should be significantly more numerous than the real data sample.
Too few random objects introduces shot noise into $\xi(r)$ measurements, especially on  small scales ($r_p<3 h^{-1}\textrm{Mpc}$).
After a set of tests, for the measurements presented in this paper we decided to use $N_R=100,000$, for which additional noise was found to be negligible

Another problem one has to face while measuring the real space correlation function is related to peculiar velocities of galaxies. 
In redshift space, these peculiar velocities distort distances computed from the redshifts $z$, and, as a result, they affect the shape of the correlation function $\xi(r)$ itself. 
The corresponding distortions are known as the coherent infall and "fingers of God" (\citeauthor{Kaiser1987} \citeyear{Kaiser1987}).
These differences, however, apply only to the radial separations and do not have any influence on the measured position of a galaxy on the sky. 
A commonly applied method to eliminate them is to split the comoving redshift space separations into two components - parallel $\pi$ and perpendicular $r_p$ to the line of sight, thus re-defining the redshift-space correlation function as $\xi(r_p,\pi)$.
Integrating $\xi(r_p,\pi)$ along the line of sight gives us a projected correlation function $w_p(r_p)$, which is the two-dimensional counterpart of the real-space correlation function, free from the redshift-space distortions (\citeauthor{DavisPeebles1983} \citeyear{DavisPeebles1983}):
  \begin{equation}
   w_p(r_p) = 2\int^{\infty}_0\xi(r_p,\pi)d\pi = 2\int^{\infty}_0\xi \left((r_p^2+y^2)^{\frac{1}{2}}\right)dy. 
\label{eq:wp}
  \end{equation}  
Here, $y$ is the real-space separation along the line of sight, and $\xi(r)$ stands for real space correlation function computed for $r=\sqrt{r_p^2 + y^2}$. 
In practice, the upper integral limit $\pi_{max}$ has to be finite in order to avoid adding noise to the result.
After performing a number of tests, we decided to use $\pi_{max} = 20$ $h^{-1} \textrm{Mpc}$, as the value around which the correlation function is the most stable.

\subsection{Error estimates}
\label{sec:errors}

Estimating errors of the two-point correlation function is the subject of numerous discussions in the literature since the time of the very first measurements (see \citeauthor{Hamilton1993} \citeyear{Hamilton1993}, \citeauthor{Fisher1994} \citeyear{Fisher1994}, \citeauthor{Bernardeau2002} \citeyear{Bernardeau2002}).
The usage of a properly constrained covariance matrix is needed to account for the fact that the values of $w_p$ for different separations $r_p$ are not independent - pair counts in different $r_p$ bins include partially the same galaxies. 
After performing a number of tests, for the VUDS data we decided to apply a combined method, which makes use of the so-called "blockwise bootstrap re-sampling" (\citeauthor{Barrow1984} \citeyear{Barrow1984}) coupled to mock catalogues (see Section \ref{sec:mock}), similar to the method proposed by \cite{Pollo2005}.

The "blockwise bootstrap re-sampling" is based on the idea of perturbing the data set by randomly creating a large number of comparable pseudo data-sets, which differ only slightly from the original sample.
A  number of objects is randomly selected from the data sample (objects are allowed to be drawn multiple times), and the correlation function is computed for each of these sub-samples.
This procedure is repeated $N_{boot}$ times, giving as a result the variation around the 'real' result, which is used as an error estimate. 

The second method makes use of sets of independent mock surveys - the simulated catalogues based on large N-body simulations coupled with physical definitions of galaxies, inserted e.g. by semi-analytic models. 
The correlation functions computed for each independent mock catalogue give the variation of the result at each scale, and are used as error estimates.

For both the bootstrap re-sampling and mocks methods the associated covariance matrix $\textrm{\textbf{C}}$ between the values of $w_p$ on $i$th and $k$th scale can be computed as:
\begin{equation}
 \textrm{\textbf{C}}_{ik} = \Big\langle\left(w_p^j(r_i) - \langle w_p^j(r_i)\rangle_j\right) \left(w_p^j(r_k) - \langle w_p^j(r_k)\rangle_j\right)\Big\rangle_j,
  \label{eq:covariance}
\end{equation}
where "$\langle\rangle$" indicates an average over all bootstrap or mock realizations, the $w_p^j(r_k)$ is the value of $w_p$ computed at $r_p = r_i$ in the cone $j$, where $1<j<N_{mock}$ for the VUDS mocks
and $1<j<N_{boot}$ for the bootstrap data.

In our case the covariance matrix reconstructed from $N_{mock}=66$ VUDS mock catalogues could not be directly applied to the observed data, because it caused the fit to be often unstable, and not able to properly converge. 
This instability was caused by the fact that the diagonal elements of the matrix have realistic values but the off-diagonal non-zero elements differ significantly from those pertaining to the data sample.
In other words, the detailed statistical properties of the VUDS mock catalogues are not close enough to the real VUDS data to allow for such an operation directly.
For this reason we computed error bars using the scatter between the VUDS mock catalogues (since we believe it is more realistic than the scatter between the bootstrap realizations), but the off-diagonal elements of the covariance matrix were computed using the classical bootstrap method. 
This means that for each redshift range we measured the correlation function $w_p(r_p)$ from 1) the VUDS data; 2) $N_{mock}=66$ VUDS mock catalogues; 3) $N_{boot}=100$ bootstrap re-samplings of the VUDS data and in the fitting and error estimation we used the combination of these measurements.

\subsection{Systematics in the CF measurement}
\label{sec:systematics}
Before introducing correlation function measurements from our data several tests have been performed on mock catalogues to investigate the influence of various VUDS survey properties on the correlation function.
As the result of these tests, the correlation function correction scheme has been performed as described in the Appendix \ref{sec:corr_correction}.

In the lower panel of Figure \ref{fig:weights_influence} it is shown that the final measurement of the correlation function $w_p(r_p)$, obtained after introducing the full correction scheme, is still slightly underestimated, by as much as $\sim 10\%$ on small scales, with respect to the 'true' correlation function obtained for the mock parent samples (see details in App. \ref{sec:corr_correction}).
This systematic underestimation may affect the value of the correlation length $r_0$, and all HOD parameters including an underestimation of the average halo mass $\langle M_h \rangle$ with respect to the 'true' underlying values.
To estimate the size of this effect, we performed a power-law function fit on the 'true' correlation function (computed for the parent mock) as well as on the 'observed' (the mock catalogue after SSPOC selection), with the full correcting scheme, both measured as the average from 66 VUDS mock catalogues.
We found that the 'true' correlation length is on average larger than the 'observed' $r_0$  by $\Delta r_0 = 0.42 \pm 0.34$,  
and that the 'observed' slope was underestimated by $\Delta \gamma = 0.06 \pm 0.04$.
This, of course, would impact the galaxy bias estimation, since it is derived directly from these parameters,
which we estimated to be underestimated by $\Delta b = 0.35 \pm 0.27$.

To address this problem we compute the ratio $w_p^{par}/w_p^{obs}$ between the 'true' $w_p^{par}(r_p)$ and 'observed' $w_p^{obs}(r_p)$ correlation function (based on the VUDS mock, see appendix \ref{sec:corr_correction}) on every scale $r_p$ for each redshift range separately.
The correlation function measurements are then multiplied by this ratio for each separation $r_p$, resulting in the final correlation function 
measurement $w_p(r_p)$ on which the power-law and HOD fitting have been performed.

All results presented in this paper include the correlation function corrections described above. 

\subsection{Power-law model}
\label{sec:pl_fitting}
In most cases, especially in the local Universe, the correlation function $\xi(r)$ is well described by a power law function $\xi(r) = (r/r_0)^{-\gamma}$, where $r_0$ and $\gamma$ are correlation length and slope, respectively.
With this parametrisation, the integral in Equation \ref{eq:wp} can be computed analytically and $w_p(r_p)$ can be expressed as:
\begin{equation}
 w_p(r_p) = r_p \Big(\frac{r_0}{r_p} \Big)^{\gamma}\frac{\Gamma(\frac{1}{2})\Gamma(\frac{\gamma-1}{2})}{\Gamma(\frac{\gamma}{2})},
\end{equation}
where $\Gamma$ is the Euler's Gamma Function.

The values of $w_p$ are not independent at different separations, hence it is not possible to use simple $\chi^2$ minimization to find the best-fit parameters.
However, the covariance matrix $\textrm{\textbf{C}}$ is symmetric and real, and therefore it can be inverted (if it is not singular).
We can fit $w_p$ by minimizing a generalized $\chi^2$, which is defined as 
\begin{equation}
\label{eq:chi2}
 \chi^2 = \sum_{i,j}^{N_D}\left( w_p^{obs}(r_i) - w_p^{mod}(r_i) \right) \textrm{\textbf{C}}^{-1}_{ij} \left( w_p^{obs}(r_j) - w_p^{mod}(r_j) \right).
\end{equation}
Here the covariance matrix $\textrm{\textbf{C}}$ is computed using the method described in section \ref{sec:errors} (Eq. \ref{eq:covariance}).
The fitting procedure to estimate the power-law parameters $r_0$ and $\gamma$ for the projected correlation function $w_p(r_p)$ follows \cite {Fisher1994} (see also \citeauthor{Guzzo1997} \citeyear{Guzzo1997} and \citeauthor{Pollo2005} \citeyear{Pollo2005}).

A more detailed description of the shape of the real-space correlation function can be done e.g. in the framework of Halo Occupation Distribution (HOD) models, which we discuss in the following section \ref{sec:hod_description}.
Nevertheless the use of a power-law model to describe the correlation function remains an efficient and simple approximation of galaxy clustering properties.

\subsection{The halo occupation distribution (HOD) model}
\begin{figure}[tb]
\centering
\includegraphics[width = .5\textwidth, height=0.5\textwidth, angle=270]{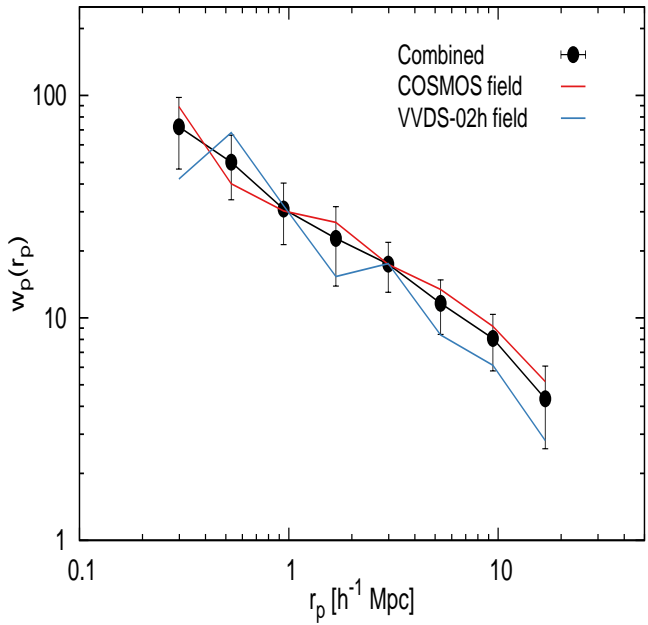}
\caption{The projected two-point correlation function $w_p(r_p)$ in each individual VUDS field for the redshift range $2<z<5$. 
	 Blue and red lines correspond to the VVDS and COSMOS field, respectively. 
	 Black points and line indicate the combined correlation function from the measurements preformed on galaxies from both fields simultaneously.}
\label{fig:cf_fields}
\end{figure}
\label{sec:hod_description}
As we mentioned above, in practice $\xi(r)$ is usually well fitted by a power law. 
However, there is no strict theoretical reason to force the galaxy correlation function to assume a power law shape. 
Indeed, recent measurements show deviations from a power law shape. 
The strongest deviations are observed on small scales, especially for the most luminous galaxies (\citeauthor{Coil2006} \citeyear{Coil2006}, \citeauthor{Pollo2006} \citeyear{Pollo2006}, \citeauthor{Zehavi2011} \citeyear{Zehavi2011}), while probably the most famous ones are baryonic acoustic oscillations (BAOs) at large scales (\citeauthor{Percival2010} \citeyear{Percival2010}).

The small scale behaviour of the correlation function of galaxies can be well interpreted in the framework of the halo occupation distribution (HOD) models based on the relations between the distribution of dark matter and galaxies.
The HOD models describe bias in terms of the probability $P(N|M)$ that a dark matter halo of a mass $M$ contains $N$ galaxies of a given type.
Recently this approach has been very successfully used to model the shape of the two-point correlation function (e.g. \citeauthor{Skibba2009} \citeyear{Skibba2009}, \citeauthor{Abbas2010} \citeyear{Abbas2010}, \citeauthor{Zehavi2011} \citeyear{Zehavi2011}, \citeauthor{Coupon2012} \citeyear{Coupon2012}, \citeauthor{Kim2014} \citeyear{Kim2014}).

In this work we apply the halo occupation model following the current commonly used analytical prescriptions, so that our results can be easily compared to results in the literature. 
We describe below the most important features and components of the HOD model we use, while for the general overview of the HOD philosophy we encourage readers to look into the review paper by \cite{SethCooray2002}. 
A very thorough description is also given by \cite{Coupon2012}.

In the HOD framework the correlation function can be split into two components. 
The one-halo component $\xi^{1h}(r)$ dominates on scales smaller than the size of dark matter haloes (usually $<1.5 \textrm{ h}^{-1} \textrm{Mpc}$), while the two-halo component $\xi^{2h}(r)$ dominates on larger scales.
The one-halo term arises from pairs of galaxies located within the same halo, while the two-halo terms is built by the pairs of galaxies hosted by different haloes.
Consequently, the correlation function can be written as:
\begin{equation}
 \xi(r) = \xi^{1h}(r) + \xi^{2h}(r).
\end{equation}
The properties of the first term, i.e. $\xi^{1h}(r)$ can be described with the use of the halo occupation model which tries to follow the distribution of galaxies located in one dark matter halo. 
The second term $\xi^{2h}(r)$, depends on the statistical properties of the large scale dark matter density field and the distribution of dark matter haloes with respect to this density field. 

The most important properties of the dark matter halo from this point of view are the halo mass function $n(M,z)$, for which we adopted the fitting formula proposed by \cite{Tinker2008}, the halo density profile $\rho(r|M)$, for which we assumed the form described by \citet*{NFW1997}, and the halo bias $b_h(M,z,r)$, for which we use the formula proposed by \cite{Tinker2010} with the scale-dependency from \cite{Tinker2005}.  

We parametrized our halo occupation model in the way used by \cite{Zehavi2005} and motivated by \cite{Kravtsov2004}.
In particular, we express the halo occupation function, i.e. the number of galaxies per halo, as a sum of a central galaxy and satellite galaxies.
The mean occupation function for central galaxies $N_c(M)$ is represented by a step function, while the mean halo occupation function for satellite galaxies $N_s(M)$ is approximated by a Poisson distribution with the mean being a power-law:
\begin{eqnarray}
 \langle N_g|M\rangle & = & 1 + \left(\frac{M}{M_1}\right)^{\alpha}\quad \textrm{for} \quad M > M_{min}\\ \nonumber
		      & = & 0	\quad \quad \quad \quad \quad \textrm{otherwise},
\end{eqnarray}
where $M_{min}$ is the minimum mass needed for a halo to host one central galaxy, and $M_1$ is the mass of a halo having on average one satellite galaxy, while $\alpha$ is the power law slope of the satellite mean occupation function.

The HOD model used in this work is simplified in comparison with the ones widely used at lower redshift ranges, e.g. by Zehavi et al. (2011) for fitting the SDSS data. 
However, based on the available statistics of the VUDS sample, with lower numbers of galaxies than e.g. in the SDSS, the three parameter model appears to be the best solution to retrieve robust measurements of $M_{min}$ and $M_1$. 
A similar approach was taken by Abbas et al. (2010) in the case of the measurements from the VVDS survey at $z \sim 1$ with a similar sample size (Abbas et al. 2010). 
Since the details of the interplay between galaxy evolution and large scale structure growth at epochs as early as the ones probed by VUDS are still poorly understood,
it is reasonable to keep a simplified description and not to dilute our analysis with models introducing too many free variables. 
The main results presented in this paper, however, are not expected to differ much from results where more complex HOD models are applied.

From the best-fit HOD parameters it is possible to obtain quantities describing halo and galaxy properties, like the average host halo mass $M_h$ 
\begin{equation}
  \label{eq:halo_mass}
 \langle M_h|g\rangle(z) = \int dM\ M\ n(M,z)\frac{\langle N_g|M\rangle}{n_{g}(z)},
\end{equation}
the large-scale galaxy bias $b_L^{HOD}$
\begin{equation}
\label{eq:hod_bias}
 b_L^{HOD}(z)=\int dM \ b_h(M) \ n(M,z)\frac{\langle N_g|M\rangle}{n_{g}(z)}
\end{equation}
where  $n(M,z)$ is the dark matter mass function, $b_h(M,z)$  is the large-scale halo bias, and the $n_{g}$ given by:
\begin{equation}
 n_g = \int n(M)\ \langle N_g|M \rangle \ dM
 \label{eq:ng}
\end{equation}
represents the number density of galaxies.

\subsection{Fitting the HOD model to the correlation function}
\label{sec:hod_fitting}
\begin{table*}[htb]
  \caption{Properties of general population VUDS samples in the redshift range $2.0<z<5.0$.}
  \begin{center}
  \scalebox{1.2}{
  \begin{tabular}{llllllll} \hline\hline
   \multicolumn{1}{l}{$z$ $\textrm{range}$} & \multicolumn{1}{l}{$z_{mean}$} & \multicolumn{1}{l}{$M_U^{median}$} & \multicolumn{1}{l}{$\log M_*^{median}$} & \multicolumn{1}{l}{$N_{gal}$} & \multicolumn{1}{l}{$r_0$} & \multicolumn{1}{l}{$\gamma$} & \multicolumn{1}{l}{$b_L^{PL}$} \\ \hline\hline
   $[2.0-5.0]$	&2.95 &	-21.55 & 9.76 & 3022 &	 $3.97^{+0.36}_{-0.38}$ &	 $1.70^{+0.09}_{-0.09}$	& $2.68^{+0.22}_{-0.21}$ \\ \hline
   $[2.0-2.9]$	&2.50 &	-21.31 & 9.66 & 1556 & 	 $3.95^{+0.48}_{-0.54}$ &	 $1.81^{+0.02}_{-0.06}$	& $2.39^{+0.32}_{-0.29}$ \\
   $[2.9-5.0]$ 	&3.47 &	-21.81 & 9.86 & 1466 & 	 $4.35^{+0.60}_{-0.76}$ &	 $1.60^{+0.12}_{-0.13}$	& $3.26^{+0.47}_{-0.37}$ \\ \hline
  \end{tabular}
  }
  \end{center}
  \label{tab:cf_redshift}
\end{table*}

\begin{figure*}
\centering
\includegraphics[height =\hsize, angle=270]{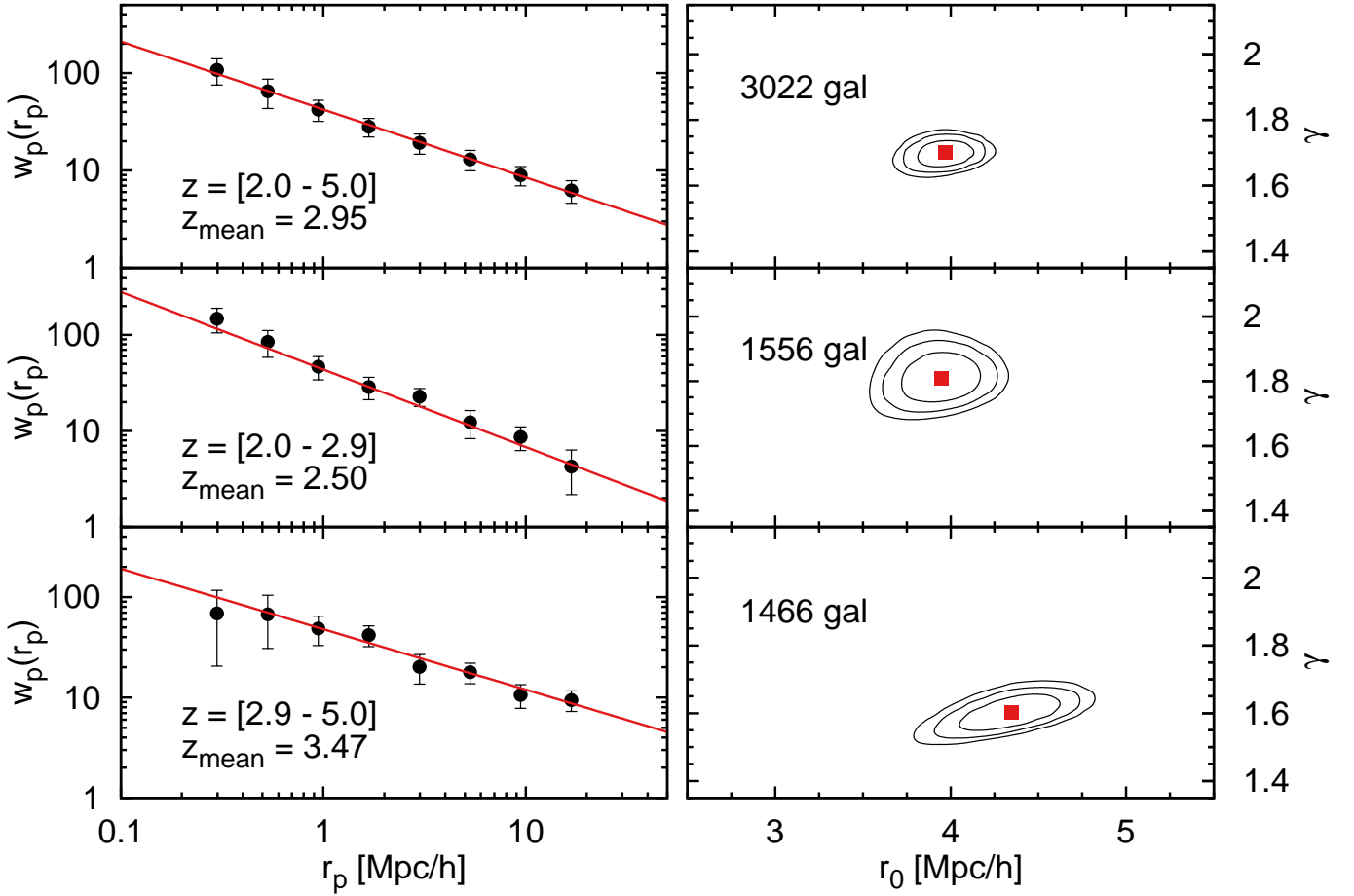}
\caption{{\it Left panel}: The projected two-point correlation function $w_p(r_p)$.  
The symbols and errorbars denote measurements of the composite correlation function from the VUDS survey. 
{\it Right panel}:  The associated best-fit power-law parameters $r_0$ and $\gamma$ along with error contours for the general galaxy population separated in three redshift ranges.}
\label{fig:cf_redshift_range}
\end{figure*}
To explore the HOD parameter space we implemented our HOD model with the publicly available CosmoPMC\footnote{\url{http://www2.iap.fr/users/kilbinge/CosmoPMC/}} code, which is using the "Population Monte Carlo" (PMC) technique to sample likelihood space (\citeauthor{Wraith2009} \citeyear{Wraith2009}, \citeauthor{Kilbinger2011} \citeyear{Kilbinger2011}).
PMC is the adaptive importance-sampling technique (\citeauthor{Cappe2007} \citeyear{Cappe2007}) which allows efficient sampling of the parameters space for the large number of samples.
For each galaxy sample we fit the projected correlation function $w_p(r_p)$ and the number density of galaxies $n_g$, by summing both contributions in log-likelihood obtained by:
\begin{eqnarray}
 \chi^2 & = & \sum_{i,j}^{N_D}\left( w_p^{obs}(r_i) - w_p^{mod}(r_i) \right) \textrm{\textbf{C}}^{-1}_{ij} \left( w_p^{obs}(r_j) - w_p^{mod}(r_j) \right) \nonumber \\
 & + & \frac{\left(n_g^{obs}-n_g^{mod}\right)^2}{\sigma^{2}_{n_g}}
\end{eqnarray}
where $n_g^{mod}$ is given by Eq. \ref{eq:ng} at the mean redshift of the sample. 
The data covariance matrix is approximated taking Eq. \ref{eq:covariance}.
The error on the galaxy number density $\sigma_{n_g}$ contains Poisson noise and cosmic variance.

\subsection{Galaxy large-scale bias}
\label{sec:bias_lin}
\begin{figure*}
 \centering
 \includegraphics[height=\hsize, angle=270]{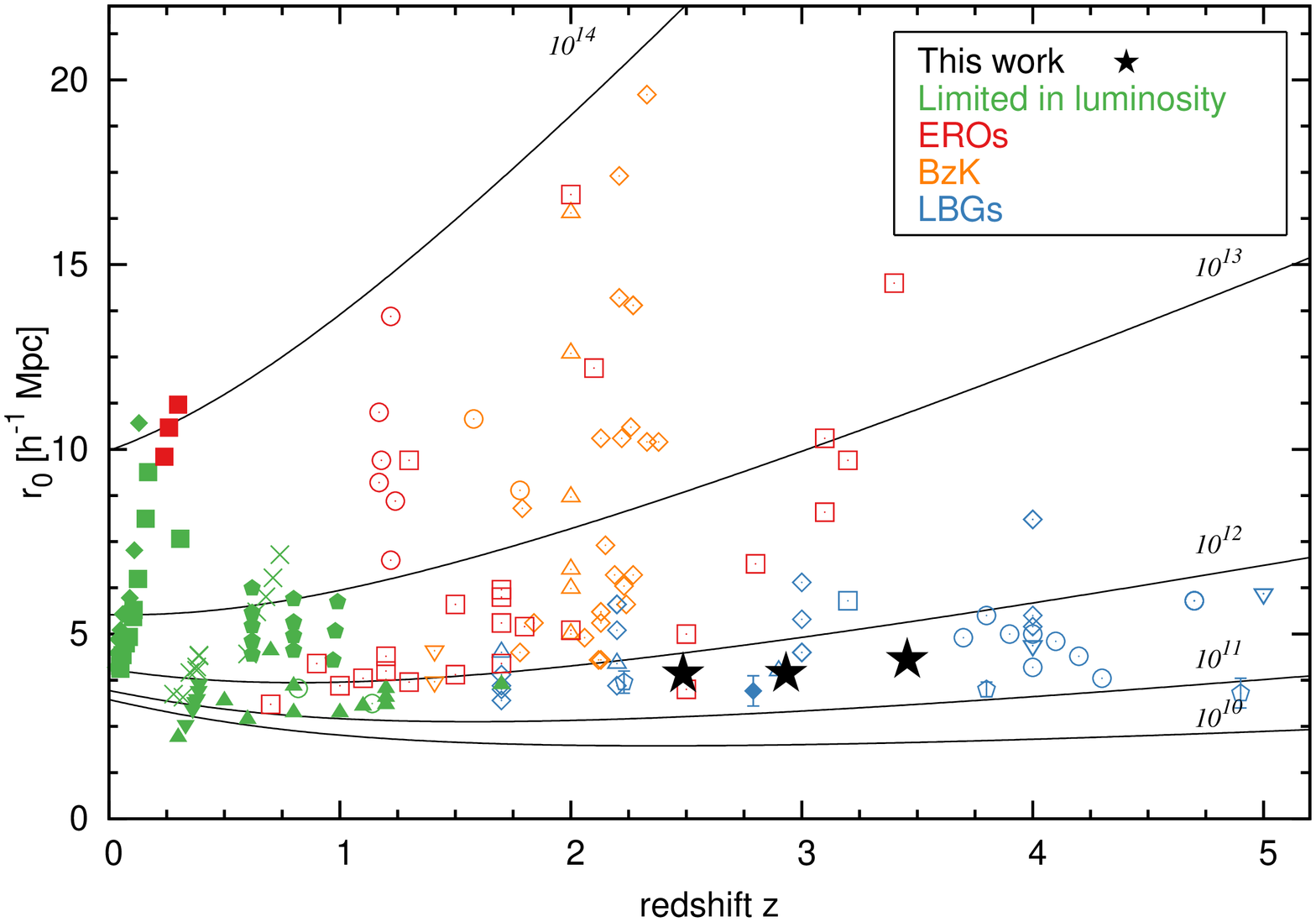}
 \caption{Correlation length $r_0$ as a function of redshift from various surveys of different types of objects. 
	  Black stars indicate the measurements obtained from this work in three redshift ranges.
	  Different colours indicate different galaxy populations targeted by using various techniques, as indicated in the upper right corner.
	  Open symbols indicate measurements based on photometric data, while filled symbols are for measurements from spectroscopic data.
	  \textit{Blue:} LBG galaxies 
	  (\citeauthor{Foucaud2003} \citeyear{Foucaud2003}, \citeauthor{Ouchi2004} \citeyear{Ouchi2004}, \citeauthor{Adelberger2005} \citeyear{Adelberger2005}, \citeauthor{Kashikawa2006} \citeyear{Kashikawa2006}, \citeauthor{Savoy2011} \citeyear{Savoy2011},  \citeauthor{Bielby2013} \citeyear{Bielby2013}, \citeauthor{Barone2014} \citeyear{Barone2014}). 
	  \textit{Orange:} B$z$K galaxies
	  (\citeauthor{Blanc2008} \citeyear{Blanc2008}, \citeauthor{Hartley2010} \citeyear{Hartley2010}, \citeauthor{McCracken2010} \citeyear{McCracken2010}, \citeauthor{Lin2012} \citeyear{Lin2012}).
	  \textit{Green:} Galaxy samples from surveys limited in luminosity
	  (\citeauthor{Norberg2002} \citeyear{Norberg2002}, \citeauthor{Coil2006} \citeyear{Coil2006}, \citeauthor{OLF2005} \citeyear{OLF2005}, \citeauthor{Pollo2006} \citeyear{Pollo2006}, \citeauthor{Zehavi2011} \citeyear{Zehavi2011}, \citeauthor{Marulli2013} \citeyear{Marulli2013}, \citeauthor{Skibba2009} \citeyear{Skibba2009}).
	  \textit{Red:} EROs or massive red galaxies
	  (\citeauthor{Daddi2003} \citeyear{Daddi2003}, \citeauthor{Zehavi2005r} \citeyear{Zehavi2005r}, \citeauthor{Brown2008} \citeyear{Brown2008}).
	  The five solid curves show the correlation length of dark halos with different minimum masses, as labelled.
	  }
	  \label{fig:r0_evolution}
\end{figure*}
According to the current cosmological paradigm of structure formation, galaxies form and evolve inside dark matter halos (\citeauthor{White1978} \citeyear{White1978}).
In other words, there exists a connection between the dark matter distribution and galaxies in the dense dark matter regions, where the halo clustering is important, galaxies should be more clustered.
The galaxy spatial distribution, however, is biased with respect to the dark matter density field.
The strength of this effect is referred to as \textit{galaxy bias}. 

As a first approximation we use a linear model for the galaxy large-scale bias. 
It assumes a linear relation between galaxy $\sigma_{R,g}(z)$ and mass $\sigma_{R,m}$ rms at a given redshift:
\begin{equation}
 \sigma_{R,g}(z) = b\sigma_{R,m}(z)
\end{equation}
where $b$ is the galaxy bias.
We also assume that $b$ is independent of scale $R$, which is true especially for large scales.
Usually the adopted value for the scale on which $\sigma$ is measured is $R = 8 h^{-1} \textrm{Mpc}$. 
Locally, at $z=0$, the mass fluctuations reach the value $\sigma_{8,m}(z=0) = 0.83$ (\citeauthor{Planck2013} \citeyear{Planck2013}).
In our model the redshift evolution of this quantity is described as:
\begin{equation}
 \sigma_{8,m}(z) = \sigma_{8,m}(z=0)D(z),
\end{equation}
where
\begin{equation}
 D(z) = \frac{g(z)}{g(0)(1+z)},
\end{equation}
and $g(z)$ is the normalized \textit{growth factor}, which describes how fast the linear perturbations grow with the scale factor.
Its value depends on the assumed cosmological model, i.e. cosmological parameters. 
After \cite{Carroll1992} we write down
\begin{equation}
 g(z) = 2.5\Omega_m\left[ \Omega^{4/7}_m - \Omega_{\Lambda} + \left(1+\frac{\Omega_m}{2}\right)\left(1+\frac{\Omega_{\Lambda}}{70}\right)\right]^{-1},
\end{equation}
where $\Omega_m$ and $\Omega_{\Lambda}$ are the matter and dark energy density parameters at a given redshift.

The galaxy rms at a given scale $R$ can be retrieved from the power-law fits to the correlation function as,
\begin{equation}
\sigma_{8,g} = \sqrt{ C_{\gamma}\left(\frac{r_0}{8 h^{-1} \textrm{Mpc}}\right)^{\gamma}},
\end{equation}
with
\begin{equation}
 C_{\gamma} = \frac{72}{2^{\gamma}(3-\gamma)(4-\gamma)(6-\gamma)}.
\end{equation}
where $r_0$ and $\gamma$ are the correlation length and the slope of power-law approximation of the correlation function.

In the following sections, we refer to the galaxy large-scale bias computed by this method as $b_L^{PL}$.
This is also used to compute the values based on measurements of the correlation function from other surveys when the $b_L$ estimations were not presented by the authors; in these cases we use their power-law best fit parameters $r_0$ and $\gamma$ (e.g. from \citeauthor{OLF2005} \citeyear{OLF2005}).
Another estimate of linear large-scale galaxy bias can be obtain from the HOD model.
We compute the  large-scale galaxy bias $b_L^{HOD}$ using the fit of the halo occupation function (Eq. \ref{eq:hod_bias}).
The comparison of these two bias values and differences between these two procedures are discussed in section \ref{sec:5_discussion}.


\section{Results}
\label{sec:4_results}
\subsection{The VUDS correlation function - general population}
\label{sec:redshift_results}
The projected two-point correlation function $w_p(r_p)$ is computed for the galaxy population with $i_{AB}<25$ at $2<z<5$ separated into three redshift ranges.
We are using a total number of 3\,022 spectroscopically confirmed VUDS galaxies located in two fields.
In Figure \ref{fig:cf_fields} we present composite correlation function $w_p(r_p)$ measurements for the full data sample in the redshift range $2 < z < 5$, along with the measurements made for each field independently.
It is worth stressing that it is the first time that the correlation function has been computed with such a high accuracy from spectroscopic data at redshift $z\sim3$.
As expected, the values of the correlation function measured for galaxies located in separate fields are slightly different due to cosmic variance.
However, for both fields the measurements are consistent with respect to the error bars and the signal is retrieved on all scales  $0.3 < r_p < 16\quad h^{-1} \textrm{Mpc}$, so that the composite function is not dominated by the signal from any of the individual fields at any separation $r_p$.

The correlation functions for the full sample z=[2,5] and the two z=[2,2.9] and z=[2.9,5] redshift intervals are presented in Figure \ref{fig:cf_redshift_range}.

\subsection{General population - Power law modelling}
We fit a power-law to $w_p(r_p)$, with two free parameters $r_0$ and $\gamma$, to quantify the clustering strength.
All the points in the range $0.3 \textrm{ h}^{-1} \textrm{Mpc}<r_p<16 \textrm{ h}^{-1} \textrm{Mpc}$ are used (see Sec. \ref{sec:pl_fitting} for the fitting method details).
The measured best-fit parameters are listed in Table \ref{tab:cf_redshift} and the power-law fit of the correlation function for each redshift sub-sample is shown in Figure \ref{fig:cf_redshift_range} together with the ($r_0$,$\gamma$) error contours.

Within our sample we find $r_0\sim3.95^{+0.48}_{-0.54}\quad h^{-1} \textrm{Mpc}$ in z=[2.0,2.9] and $r_0\sim4.35^{+0.60}_{-0.76}\quad h^{-1} \textrm{Mpc}$ in z=[2.9,5].
The slight increase of $r_0$ with redshift is probably due to the small luminosity differences between samples (see Sec. \ref{sec:5_discussion} for discussion), although the difference is marginally significant given measurement errors. 
For these samples, the slope $\gamma$ varies between $\gamma=1.60^{+0.12}_{-0.13}$ and $\gamma=1.81^{+0.02}_{-0.06}$, hence showing a tendency to decrease with redshift. 

\subsection{General population - HOD modelling}

The best-fit HOD model defined in section \ref{sec:hod_description} is fit to each of three redshift sub-samples as presented in the left panels of Figure \ref{fig:hod_fit_redshift}. 
The HOD model provides  a good fit to the real-space correlation function $w_p(r_p)$ in all the redshift ranges, particularly in z=[2,2.9] where it is an excellent representation of the data. 
For the highest redshift bin z=[2.9,5] the fit is not as good. 
This is mainly due to the more noisy measurement of the correlation function in this redshift range.

The three HOD parameters, the minimum halo mass $M_{min}$, the satellite occupation halo mass $M_1$, and the high halo mass slope $\alpha$ (see sec. \ref{sec:hod_description}), are inferred from the full covariance matrix (see section \ref{sec:hod_fitting}) and are listed in Table \ref{tab:hod_params}.
The parameters are allowed to vary within a large range set from previous observations at high and low redshift: $10 < \log M_{min} < 14$, $10 < \log M_1 < 15$, $0.6 < \alpha < 2.0$.
The right panel of Figure \ref{fig:hod_fit_redshift} presents the halo occupation function obtained from the HOD fit within three redshift ranges.
We do not observe any significant difference in halo masses between the two redshift ranges probed for the general VUDS galaxy population.
The minimum halo mass, for which a halo hosts at least one central galaxy $M_{min}$, has comparable value for the $z\sim2.5$ and $z\sim3.5$ samples.

The satellite occupation mass $M_1$ is noticeably higher in the higher redshift bin compared to the lower redshift sample, although the errors are quite large.
These uncertainties are related to the weak one-halo term signal in the correlation function for the higher redshift measurement.

The high halo mass slope $\alpha$ of the satellite mean occupation function is around $\sim1.3$ for z=[2,2.9], significantly steeper than $\alpha=0.73$ found in z=[2.9,5].

The average host halo mass $<M_h|g>$, large-scale galaxy bias $b_L^{HOD}$ and associated $1\sigma$ errors, are reported in Table \ref{tab:hod_params}.
The redshift evolution of these parameters is presented in Figure \ref{fig:eff_mass_vdb_zoom} - for the average host halo mass, and in Figure \ref{fig:bias} - for the large-scale galaxy bias.
These results are fully discussed in Section \ref{sec:5_discussion}.

\begin{table*}[t]
  \caption{The HOD and and HOD-based parameters for the redshift sub-samples}
  \begin{center}
  \scalebox{1.2}{
  \begin{tabular}{llllll} \hline\hline
   \multicolumn{1}{l}{$z$ $range$} & \multicolumn{1}{l}{$\log M_{min}$} & \multicolumn{1}{l}{$\log M_1$} & \multicolumn{1}{l}{$\alpha$} & \multicolumn{1}{l}{$\log \langle M_h \rangle$} & \multicolumn{1}{l}{$b_L^{HOD}$}\\ \hline\hline
    $[2.0-5.0]$		&	$11.04^{+0.33}_{-0.36}$		&	$12.11^{+0.51}_{-0.45}$		&	$1.29^{+0.12}_{-0.11}$		&	$11.75^{+0.23}_{-0.28}$		&	$2.82^{+0.27}_{-0.16}$		\\ \hline
    $[2.0-2.9]$		&	$11.12^{+0.23}_{-0.35}$		&	$12.09^{+0.32}_{-0.35}$		&	$1.30^{+0.08}_{-0.10}$ 		&	$12.01^{+0.16}_{-0.18}$		&	$2.55^{+0.16}_{-0.21}$		\\
    $[2.9-5.0]$		&	$11.18^{+0.56}_{-0.70}$		&	$12.55^{+0.85}_{-0.88}$		&	$0.73^{+0.23}_{-0.30}$ 		&	$11.61^{+0.46}_{-0.40}$		&	$3.48^{+0.31}_{-0.26}$\\ \hline
 \end{tabular}
  }
  \end{center}
  \label{tab:hod_params}
\end{table*}

\begin{figure*}
\centering
 \includegraphics[height=\hsize, angle=270]{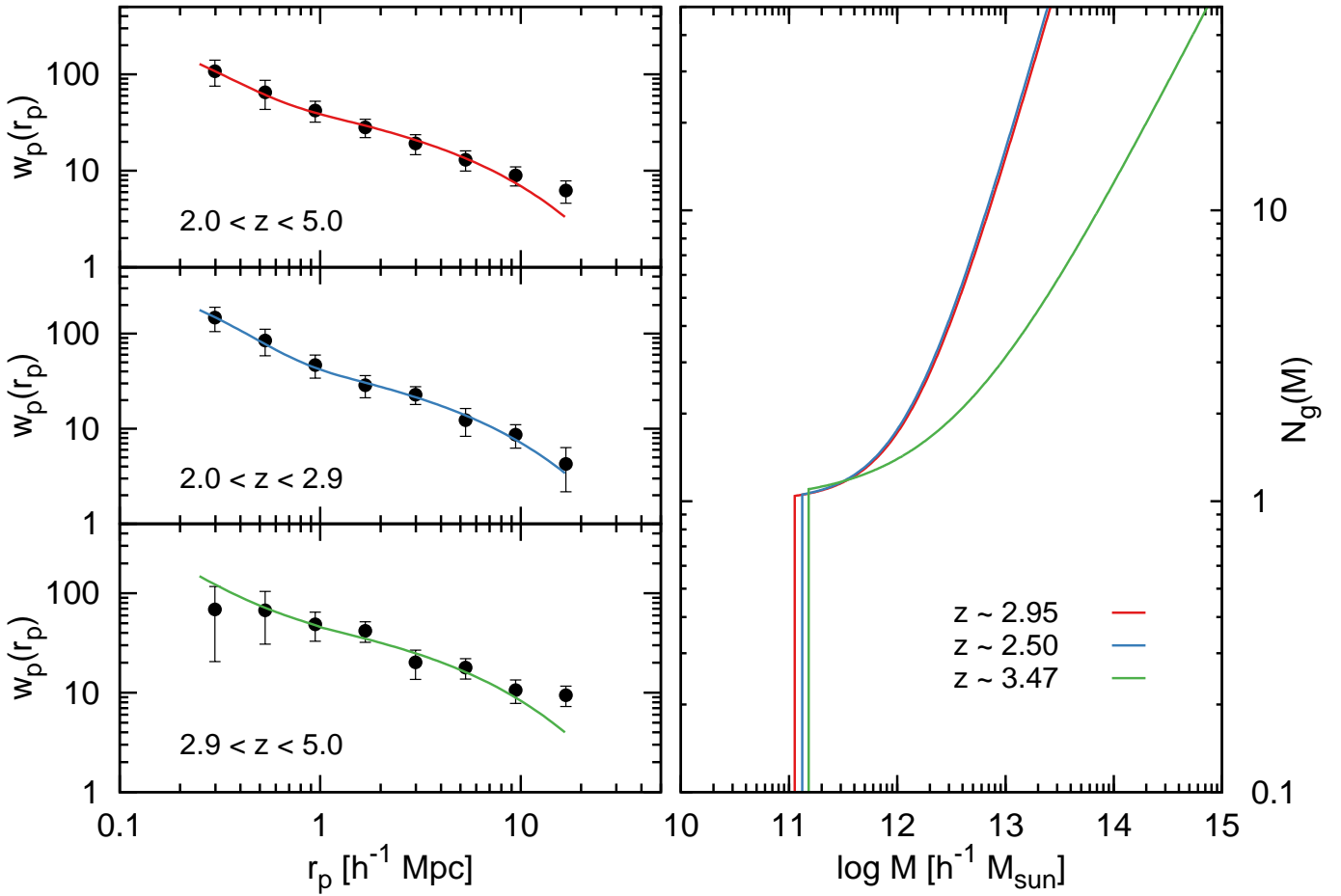}
 \caption{\textit{Left panel:} The correlation function for three redshift sub-samples as in Figure \ref{fig:cf_redshift_range}.  
	  The solid lines present the best-fit halo occupation model.
	  \textit{Right panel:} The evolution of the halo occupation for the three redshift ranges.}
	  \label{fig:hod_fit_redshift}
\end{figure*}

\subsection{Linear large-scale bias measurements}
The large-scale linear galaxy bias $b_L^{PL}$ has been computed using the method described in Sec. \ref{sec:bias_lin}, with the best-fit correlation parameters obtained for each galaxy sub-sample. 
The bias values obtained and the associated $1\sigma$ errors are listed in Table \ref{tab:cf_redshift}.
We find that the galaxy bias is increasing from $2.39\pm0.32$ at  $z\sim2.5$ to $3.26\pm0.47$ at $z\sim3.5$, galaxies in the higher redshift sample appearing to be more biased than those in the lower redshift sample.
Additionally, we note that the bias reaches a relatively high value at $z\sim3.5$.
This is further discussed in Sec. \ref{sec:5_discussion}.

\section{Discussion}
\label{sec:5_discussion}

\subsection{Evolution of the clustering length $r_0$ from $z\sim3.5$ until the present epoch}
In this paragraph we discuss the evolution of clustering properties of the general galaxy population since a redshift $z\sim3.5$ (see Sec. \ref{sec:4_results}) in the view of other measurements presented in the literature. 

Within our sample, we observe the clustering strength to increase slightly with redshift over the range $z=[2.0-5.0]$, with $r_0$ being the highest at the highest redshift $z\sim3.5$.
However, this increase is marginally significant given the measurement errors.
The observed increase of $r_0$ with $z$, surprising at the first glance, may be caused by the luminosity difference between two samples at $z\sim2.5$ and $z\sim3.5$. 
Due to the the magnitude selection $i_{AB} \leq 25$ of the VUDS sample we observe systematically more luminous galaxies at higher $z$. 
As a result, our high redshift sample contains on average more luminous galaxies, with median $M_U = -21.81$, while for the lower redshift sample the median $M_U = -21.31$. 
The luminosity dependence of galaxy clustering is a well established fact; it has been observed both at early and late cosmic epochs
(e.g. \citeauthor{Zehavi2011} \citeyear{Zehavi2011}, \citeauthor{Coupon2012} \citeyear{Coupon2012}, \citeauthor{Abbas2010} \citeyear{Abbas2010}, \citeauthor{Pollo2006} \citeyear{Pollo2006})
and luminous galaxies at all $z$ investigated so far were found to be more strongly clustered than the fainter ones which is usually related to the fact that they are located in more massive dark matter haloes and, consequently, stronger fluctuations of the dark matter density field.
If this trend is also true for the redshift range investigated in this work, the observed increase of  $r_0$ with $z$ may be caused by the fact that our higher redshift sample is on average more luminous that the lower one.
The dependency of clustering on luminosity at these high redshifts will be investigated in a future paper (Durkalec et al. in prep).

The correlation power-law parameters  $r_0=3.97\pm0.48$ and $\gamma=1.70\pm0.09$ at $z\sim3$ are comparable with the clustering parameters for the star forming galaxies at intermediate ($z\sim1$) redshift 
(e.g. \citeauthor{OLF2005} \citeyear{OLF2005}, \citeauthor{Pollo2006} \citeyear{Pollo2006}), 
but the clustering length is lower than  found for  galaxies in the local Universe 
(e.g. \citeauthor{Zehavi2011} \citeyear{Zehavi2011}, \citeauthor{Norberg2002} \citeyear{Norberg2002}). 
This has to be placed in the global perspective of the clustering expected to grow stronger along cosmic time. 

A direct comparison of clustering strength at different epochs is a difficult task for several reasons.
Different measurements are based on various galaxy samples, each with a unique selection function. 
Since galaxies cluster differently depending on their properties (like luminosity, colours or stellar masses), and in practice every catalogue contains a galaxy population somewhat different, the differences in clustering measurements may primarily be attributed to different galaxy sample properties as much as to evolutionary effects.
The inhomogeneities between galaxy samples are especially pronounced for the high redshift samples. 
At $z>2$ correlation functions have mostly been measured from surveys targeting specific classes of galaxies, e.g. extremely massive red objects or sources selected using a Lyman-break technique selecting star-forming galaxies.
In addition, until recently most clustering measurements at $z>2$ were produced from the angular correlation function computed on photometric samples with the knowledge of the redshift being only approximate and based on color selection 
(BzK e.g. \citeauthor{Daddi2003} \citeyear{Daddi2003}, \citeauthor{McCracken2010} \citeyear{McCracken2010}; LBG e.g. \citeauthor{Adelberger2005} \citeyear{Adelberger2005}; LAE e.g. \citeauthor{Ouchi2004} \citeyear{Ouchi2004}). 
Compared to spectroscopic samples, the lower accuracy and precision in redshift measurements in these samples and contamination in a given redshift bin from galaxies at other redshifts introduces a blurring of large scale structures which is difficult to overcome even when using large samples.

Our $r_0$ estimations are in excellent agreement with other available results at similar high redshifts 
(e.g. \citeauthor{Daddi2003} \citeyear{Daddi2003}, \citeauthor{Ouchi2004} \citeyear{Ouchi2004}, \citeauthor{Adelberger2005} \citeyear{Adelberger2005}, \citeauthor{Kashikawa2006} \citeyear{Kashikawa2006}, \citeauthor{Hartley2010} \citeyear{Hartley2010}, \citeauthor{Savoy2011} \citeyear{Savoy2011}), 
as presented in Figure \ref{fig:r0_evolution}. While $r_0$ in the VUDS sample is on the lower side of previous clustering measurements
at these redshifts, most other samples have specific selection functions from which is 
it expected that they have a higher clustering bias than our sample, particularly for the redder or more luminous samples.
Our measurements can be compared to the measurements by \cite{Bielby2013}, based on a large sample of $2135$ spectroscopically confirmed LBGs in $\sim10\,000$ arcmin$^2$ at $z=2.8$, the largest sample with a spectroscopically-based measurement from a large volume at these redshift prior to our measurements. 
They find a clustering length  $r_0 = 3.46 \pm 0.41$ at $z=2.79$ and a slope $\gamma=1.5-1.6$, flatter than for local galaxies, in good agreement with our measurements given the associated errors.
\cite{Savoy2011} measured a correlation length $r_0 = 3.5\pm0.5$ from $\sim1400$ UV-selected galaxies  with photometric redshifts $z\sim3$ 
and $23 < R < 27$ in the Keck Deep Fields, and \cite{Geach2012} found  $r_0 = 3.7\pm0.3$ from the clustering of 
$\textrm{H}\alpha$ emitters (HAEs) at $z=2.23$ from HiZELS, both results being very similar to ours at comparable redshifts.
We conclude that our measurements are in general agreement with the most recent and least biased measurements available in the literature.
Our results are among the most reliable today given the VUDS survey area and the number of galaxies used in our study.

\subsection{The host halo mass}
\label{sec:halo_mass_discussion}
\begin{figure*}
\centering
\includegraphics[height = \hsize, angle =270]{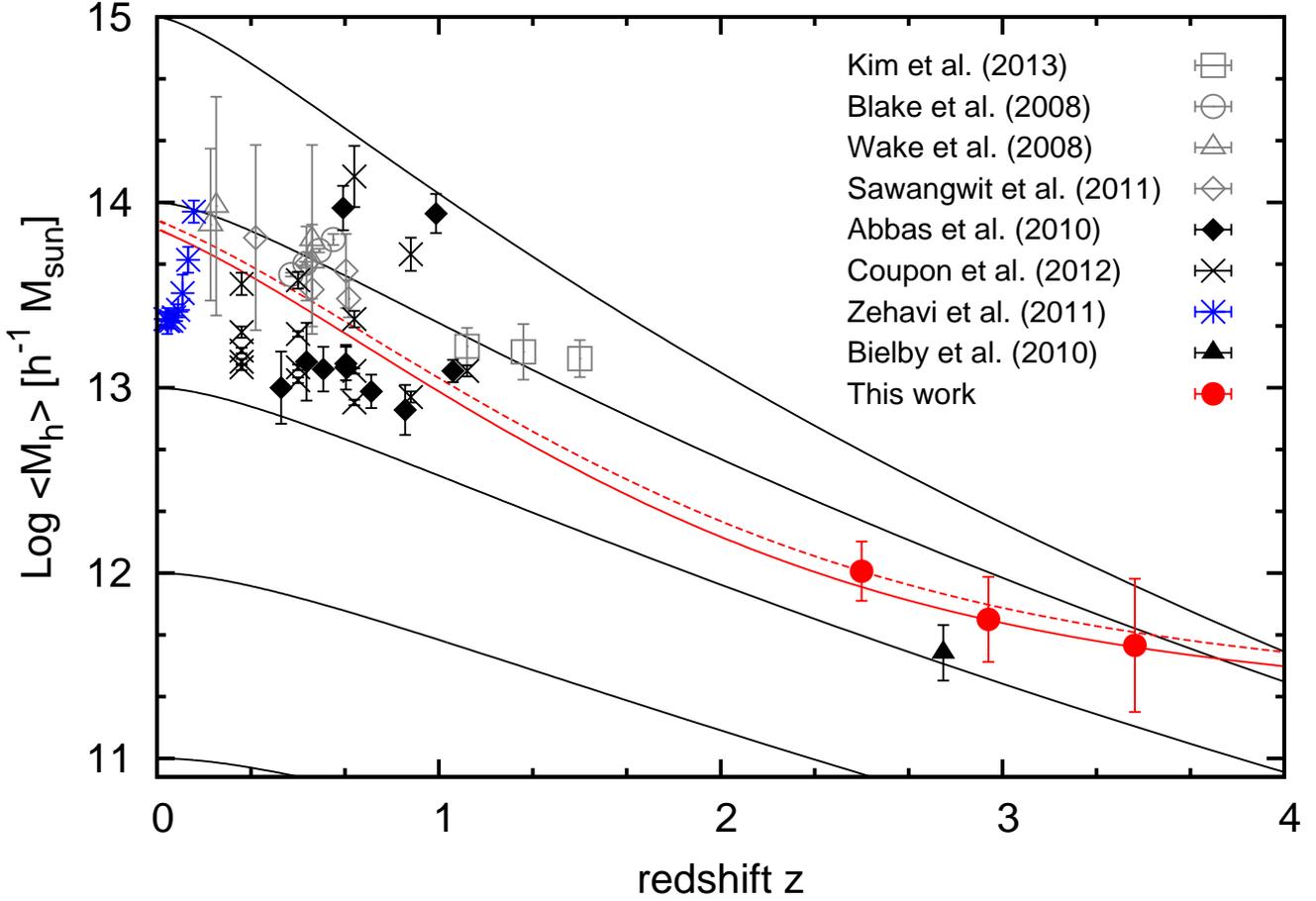}
\caption{The evolution of the number-weighted average host halo mass given by Eq. \ref{eq:halo_mass} for the three redshift ranges analysed in this study.
	The red filled circles indicate mass estimations from VUDS. 
	Black and grey symbols represent the results of previous work based on spectroscopic and photometric surveys respectively.
	The solid black lines indicate how a host halo of a given mass $M_0$ at $z=0$ evolves with redshift, according to the model given by \cite{vandenBosch2002}. 
	The solid red line represents the halo mass evolution derived using Eq. \ref{eq:mass_growth_my}, with the HOD parameters obtained from the best-fit HOD model at a redshift $z\sim3$.
	The dashed red line is using the HOD best-fit parameters for $z\sim2.5$. 
        VUDS galaxies with a typical L$_{\star}$ luminosity are likely to evolve into galaxies with a luminosity $>$L$_{\star}$ today.}
\label{fig:eff_mass_vdb_zoom}
\end{figure*}

We draw in Figure \ref{fig:r0_evolution}  the model predictions for the evolution of $r_0$ of dark matter halos following the prescription given by \cite{Mo2002} (which is based on the \cite{Press1974} formalism).
According to this model, dark matter halos of an average mass $\langle M_h \rangle \sim 10^{11.5} M_{\odot}$ have the same $r_0$ as observed VUDS galaxies at $z\sim3$.
If we take the rough assumption that at $z\sim3$ haloes are occupied mostly by only one galaxy (central), this  mass would be the average mass of a halo hosting the typical VUDS galaxy.   
However, this result is only a first approximation, since it is based on simplified assumptions of the \cite{Press1974} theory and needs to be refined.

The halo occupation model (HOD) is expected to give more accurate predictions, since it accounts for the actual number and distribution of galaxies occupying haloes.
We obtain the number weighted average mass $\langle M_h \rangle$ of haloes hosting VUDS galaxies using the best-fit HOD parameters (Eq. \ref{eq:halo_mass}), in three redshift ranges.
We find that the halo mass decreases with redshift (see Figure \ref{fig:eff_mass_vdb_zoom}), evolving from $\log \langle M_h \rangle = 11.61_{-0.40}^{+0.46} h^{-1} M_{\odot}$ at $z\sim3.5$ to $\log \langle M_h \rangle = 12.01^{+0.16}_{-0.18} h^{-1} M_{\odot}$ at $z\sim2.5$. 
These estimations are in reasonable agreement with the recent observations by \cite{Bielby2013} who found 
an average mass of $\log \langle M_h \rangle \sim 11.57\pm0.15 M_{\odot}$ for haloes hosting LBG galaxies at a redshift $z\sim 2.79$.

It can also be seen from Fig. \ref{fig:eff_mass_vdb_zoom} that our estimated host halo masses are much lower than the ones typically observed in the local Universe. 
At $z\sim3$ the typical host halo masses are located in the range between $\langle M_h \rangle \sim 10^{11} - 10^{12} M_{\odot}$ while locally these masses usually reach the values between $10^{13} - 10^{14} M_{\odot}$.
These results indicate that at some stage dark matter halos must have experienced a rapid accretion phase, as expected
in the framework of the hierarchical mass assembly.

A universal formula for the mass growth of cold dark matter halos has been derived by \cite{vandenBosch2002}. 
The history of a halo with a given mass $M_0$ at $z=0$ can be traced back in time using a simple formula:
\begin{equation}
 \log\langle \Psi \left(M_0,z\right)\rangle = -0.301 \left[ \frac{\log\left(1+z\right)}{\log\left(1+z_f\right)}\right]^{\nu},
 \label{eq:mass_growth_vdb}
\end{equation}
where $z_f$ and $\nu$ are free fitting parameters that depend on halo mass and cosmological parameters. 
To obtain these fitting parameters we follow the analytical formula from \cite{vandenBosch2002}.
We then trace the evolution of the galaxy population sampled by VUDS from $z\sim3$ to the present epoch to predict the mass of halos hosting the 
present day descendants of the VUDS galaxies.
We find that in this model the typical VUDS halo with a mass $\langle M_H\rangle \sim 10^{11.75} M_{\odot}$ at $z\sim3$ should evolve into 
a halo with a mass $\langle M_H\rangle \sim 10^{13.5} M_{\odot}$ at $z=0$.
In the local SDSS galaxy sample \cite{Zehavi2011} found that halos with these masses are typically occupied by star forming galaxies with luminosity $M_r < -20.5$. According to the \cite{vandenBosch2002} model the star forming galaxies at $z\sim3$ in VUDS with a typical characteristic luminosity L$_{\star}$ would likely evolve into galaxies equivalent to or brighter than L$_{\star}$ at the present day. 

\begin{figure*}[t]
 \centering
 \includegraphics[height = \hsize, angle = 270]{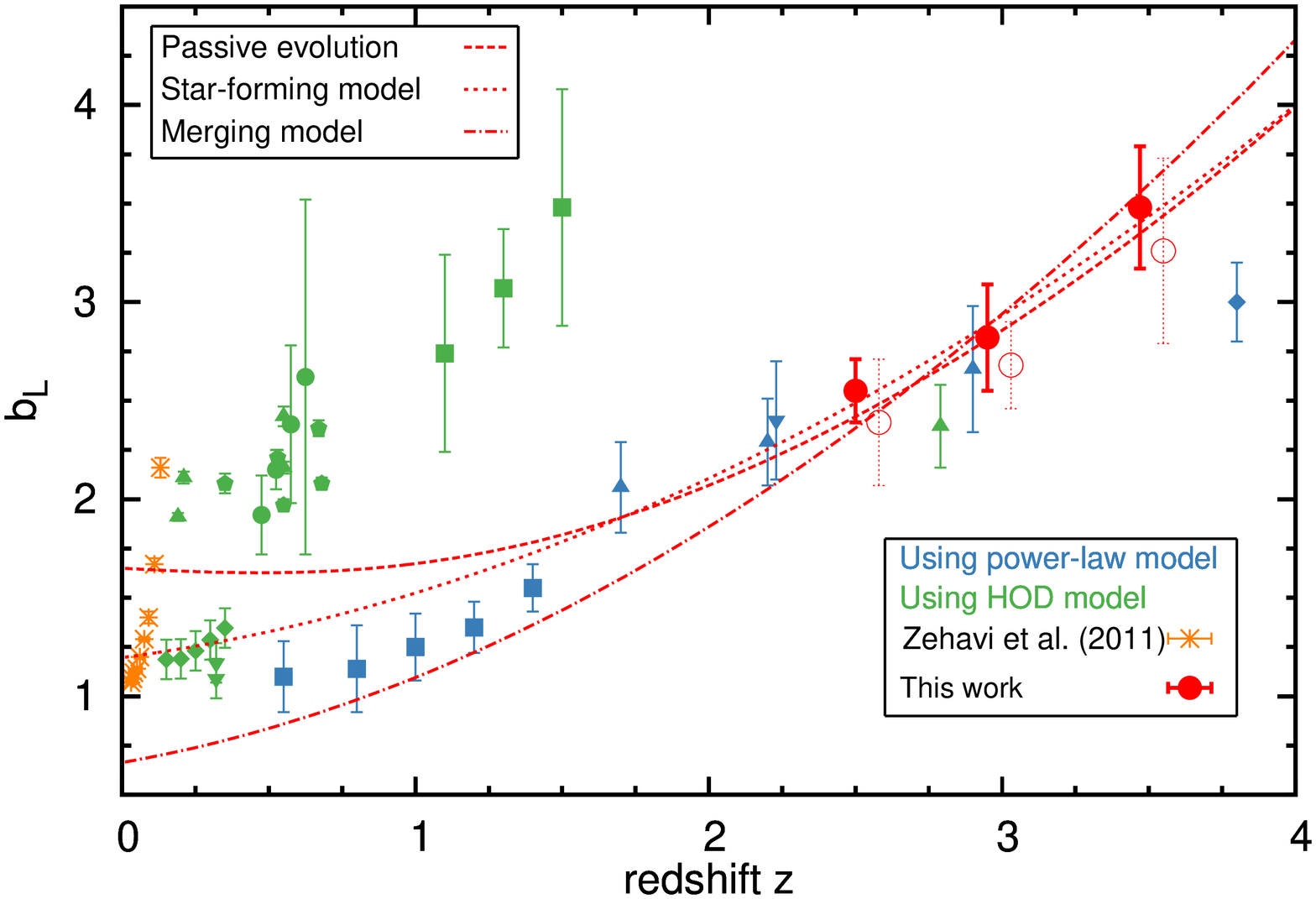}
 \caption{Large-scale linear galaxy bias retrieved by using two methods. 
	  Red filled circles indicate measurements of $b_L^{HOD}$, while the open red circles represent $b_L^{PL}$) in three redshift ranges.
          Green and blue point represents the results from previous works, retrieved by using HOD 
          (\citeauthor{Blake2008} \citeyear{Blake2008}, \citeauthor{Wake2008} \citeyear{Wake2008}, \citeauthor{Ross2009} \citeyear{Ross2009}, \citeauthor{Sawangwit2011} \citeyear{Sawangwit2011}, \citeauthor{Kim2014} \citeyear{Kim2014}) 
          and power-law frameworks (\citeauthor{Marinoni2005} \citeyear{Marinoni2005}, \citeauthor{Adelberger2005} \citeyear{Adelberger2005}, \citeauthor{Geach2012} \citeyear{Geach2012}, \citeauthor{Barone2014} \citeyear{Barone2014}, \citeauthor{Bielby2013} \citeyear{Bielby2013}) 
          We show 3 different models for the evolution of bias: a \textit{passive} evolution model (red dashed line)
          a \textit{merging} model (red dot-dashed line), a \textit{star-forming} model (dotted line), 
          as described in the Sec. \ref{sec:bias_dis}.}
 \label{fig:bias}
\end{figure*}

The above comparison assumes that each halo is occupied by only one galaxy. This is not expected to generally be the case, and the above picture, while broadly correct might need to be adjusted.
In order to trace the evolution of dark matter haloes and the hosted galaxy population in a more realistic way 
we use both the halo mass growth model $\Psi(M_0,z)$ and the halo occupation function $\langle N_g|M \rangle$ at redshift $z=3$.
The average halo mass $M_h$ as a function of redshift $z$ is measured taking (see Eq. \ref{eq:halo_mass} for comparison):
\begin{equation}
 \langle M_h \rangle (z) = \int dM \Psi^{-1}(M,z) n(M,z) \frac{\langle N_g|M\rangle}{n_{g}(z)},
 \label{eq:mass_growth_my}
\end{equation}
here the $\Psi^{-1}(M,z)$ is the inverse mass growth function proposed by \cite{vandenBosch2002} (Eq. \ref{eq:mass_growth_vdb}), $n(M,z)$ is the dark matter mass function, and $n_{g}$ the galaxy number density.
This allows us to trace the history of the typical dark matter halo hosting the average VUDS galaxy from a redshift $z=3$ to the present day, as presented in Figure \ref{fig:eff_mass_vdb_zoom}.
We find that the typical halo of mass $\langle M_h \rangle = 10^{11.75}$ at $z\sim3$ should evolve into a halo of mass $\langle M_h \rangle = 10^{13.9}$ at redshift $z=0$.
Comparing again our results from this improved model with \cite{Zehavi2011}, halos with these masses are typically occupied by star forming galaxies with $M_r<-21.5$, above the characteristic luminosity M$_{\star}$ at $z\sim0$. 
This means that the general galaxy population observed at redshift $z=3$ evolves into the brightest and most massive galaxies, 
which occupy the most massive halos observed in the local Universe, as is expected in the hierarchical mass growth paradigm.

\subsection{Galaxy bias}
\label{sec:bias_dis}

The hierarchical model of galaxy formation predicts an increase of the galaxy bias with redshift \cite{MoWhite1996}.
In this section we discuss the relation between the distribution of galaxies and the underlying dark matter density field at different 
redshifts, by comparing the galaxy bias values measured from VUDS to the bias of galaxy populations at different redshifts from the literature.

Figure \ref{fig:bias} shows the linear large-scale galaxy bias $b_L$, computed  
using methods based on power-law $b_L^{PL}$ (see Sec. \ref{sec:bias_lin}) and HOD $b_L^{HOD}$ approximations (see Sec. \ref{sec:hod_description}).
The results obtained from both methods are comparable within errors, and small differences are likely due to the less accurate power-law approximation. 
The general VUDS galaxy population shows a bias decreasing with decreasing redshift.
According to our measurements, from $z\sim3.5$ to $z\sim2.5$ the linear bias decreased by $\Delta b_L\sim0.9$. Both our values at $z=2.5$
and $z=3.5$ are significantly higher than measurements at later epochs reported in the literature.
For comparison, we plot the linear large--scale bias inferred for the intermediate ($z=[0.5-1.5]$) redshift range by \cite{Marinoni2005}, based on VVDS-Deep survey,  for $z>2$ by \cite{Geach2012} based on observations of $H\alpha$ emitters, and by \cite{Barone2014} for $z\sim3.8$ $LBGs$, along with the values we computed using the best-fit power-law parameters $r_0$ and $\gamma$ from \cite{OLF2005}, \cite{Adelberger2005} and \cite{Bielby2013} (the method is described in section \ref{sec:bias_lin}).
The values presented in Table \ref{tab:hod_params} are similar to the bias measurements recently observed for $LBGs$ at redshift $z>2$.
\cite{Bielby2013} found $b_{gal} = 2.33\pm0.17$ from their LBG-selected spectroscopic sample, and  at an even higher redshift $z\sim3.8$ \cite{Barone2014} estimated $b_L = 3.0\pm0.2$ for similarly selected galaxies.

The visible decrease of the galaxy bias with cosmic time can be explained in terms of the hierarchical scenario of structure formation.
In this framework galaxies formed at early epochs in the most dense regions (highest peaks of the density field) 
which are the most biased with respect to the underlying average mass density field.
As the mass density field evolved with time, galaxy formation is then expected to systematically move into less dense, hence less biased, regions.
Additionally, it is likely that the gas in the most dense regions became too hot for collapse, and thus star formation could not take place in the strongest over-densities any more (e.g. \citeauthor{Blanton1999} \citeyear{Blanton1999}).
Therefore, the observed decrease of galaxy bias with cosmic time may be explained as galaxy formation is moving to 
less dense areas as cosmic time increases in a downsizing trend (\citeauthor{deLucia2006} \citeyear{deLucia2006}).

In Figure \ref{fig:bias} we also draw the predicted evolution of the large-scale galaxy bias using different prescriptions. 
We consider three different theoretical descriptions of the biasing functions, based on different ideas of how evolution may proceed: the \textit{passive}, the \textit{merging}, and the \textit{star forming} biasing models (see e.g. \citeauthor{Marinoni2005} \citeyear{Marinoni2005}).
In the first model the number of galaxies (halo occupation) is conserved as a function of time and the bias evolves assuming the form 
presented in Eq. \ref{eq:hod_bias}.
The second model takes into account galaxy merging, as proposed by \cite{MoWhite1996}, who gave analytical prescriptions for computing the bias of halos using the Press \& Schechter formalism.
If we assume that galaxies can be identified with dark matter halos, an approximate expression for the biasing of all halos of mass $> M$ existing at redshift $z$ (but which collapsed at redshift greater than the observation redshift, see discussion in \citeauthor{Matarrese1997} \citeyear{Matarrese1997}) is given by:
\begin{equation}
 b(M,z) = 1 + \frac{1}{\delta_c} \left(\frac{\delta_c^2}{\sigma^2(M,z)} -1  \right)
\end{equation}
where $\delta_c \simeq1.69$ is the linear overdensity of a sphere which collapses in an Einstein-de Sitter Universe and $\sigma(M,z)$ is the linear rms fluctuations on scales corresponding to mass $M$ at the redshift of observation.
The \textit{star-forming} model assumes that the distribution of galaxies with luminosity $> L$ is well traced by halos with mass $> M$, and predicts the biasing of objects that just collapsed at the redshift of observation (e.g., \citeauthor{Blanton2000} \citeyear{Blanton2000}). 
In this model,
\begin{equation}
 b(M,z) = 1+\frac{\delta_c}{\sigma^2(M,z)}
\end{equation}
represents the biasing of galaxies that formed in a narrow time interval around redshift $z$ (i.e., galaxies which experienced recent star formation at redshift $z$).

From the halo mass analysis presented in Section \ref{sec:halo_mass_discussion} we infer that VUDS galaxies would evolve into galaxies with a luminosity $M_r<-21.5$ in the local universe (Section \ref{sec:halo_mass_discussion}), for which the bias is $b_L=1.7$. 
The star forming model predicts that the VUDS  galaxy population would evolve into a weakly biased class of galaxies, corresponding to faint low mass
galaxies.
The merging model seem to have an evolution too steep compared to what is observed as it predicts an evolution to a bias significantly lower than populations of galaxies with the lowest clustering at z=0 \citep{Zehavi2011}. 
The model that would make the VUDS galaxies with $b_L\sim2.8$ at z$=$3 evolve to $b_L=1.7$ at z$\sim$0 is the passive model.
However, it is likely that more sophisticated models would need to be developed to reproduce the joint evolution
of DM halo mass and galaxy bias.

\section{Summary}
\label{sec:6_summary}
We examined the galaxy clustering properties of high redshift galaxies with 2$<$z$<$5, by measuring the two-point real-space correlation function $w_p(r_p)$ for 3022 galaxies from the VUDS survey.
These measurements are among the first performed from spectroscopically confirmed galaxies at such high redshifts and allow to probe
clustering evolution and galaxy host halos since very early times.
We quantify our observations in terms of a power-law approximation and a three-parameter HOD model fitted to the correlation function computed in three redshift ranges, with median $z\sim2.5$, $z\sim3.0$ and $z\sim3.5$.
Our work therefore complements and extends to higher redshifts previous galaxy clustering analyses.

The results and conclusions can be summarized as follows:
\begin{itemize}
 \item We observe a mild clustering evolution for the star-forming population in the VUDS survey.
The correlation length increases from $r_0=3.95\pm 0.48$ at $z\sim2.5$ to reach $r_0=4.35\pm0.60$ at $z=3.5$.
We attribute this slight increase to the difference in absolute magnitudes between these two samples, the higher 
redshift sample being more luminous. The slope of the correlation function
is found to be $\gamma=1.81^{+0.02}_{-0.06}$ and  $1.60^{+0.12}_{-0.13}$ at $z=2.5$ and $3.5$ respectively, similar or slightly flatter than the canonical $\gamma=1.8$ observed at low redshifts.
These measurements are in general agreement with the most recent and least biased measurements available in the literature at similar redshifts. 
We find that the power-law parameters we derive from the correlation function measurements are comparable to the clustering parameters for 
star-forming galaxies at intermediate redshifts, but the clustering length is somewhat lower than for galaxies in the local Universe.      
 \item The average halo mass obtained from our HOD model fit reaches $\langle M_h\rangle=10^{11.75} h^{-1} M_{\odot}$ at redshift $z=3$,
which is smaller than observed locally.
Assuming the mass growth model from \cite{vandenBosch2002} these halos should evolve into halos with a mass $\langle M_h\rangle (z=0) = 10^{13.9} h^{-1} M_{\odot}$ at $z=0$. 
As halos with such a mass are occupied by star forming galaxies with $M_r<-21.5$ in the local universe.
we infer that the star-forming galaxy population observed at a redshift $z\sim3$ has evolved into the massive and bright galaxy 
population observed today, as predicted in the hierarchical mass growth paradigm.
 \item We found that the linear large-scale galaxy bias $b_L$, computed using both the HOD and power-law best-fit parameters, increases with redshift from $b_L^{PL} = 2.68\pm0.22$ at $z=2.5$ to $b_L^{PL} = 3.26\pm0.47$ at $z=3.5$.
These bias values are comparable with measurements obtained at similar redshifts but are much higher than bias values observed at low and 
intermediate redshift ranges where $b_L$ reaches values in the range $1-2$.
We discuss simple models to link the z$\sim$3 galaxies in our sample to lower redshift populations,
and find that a passive halo evolution model maintaining halo occupation parameters constant will lead to a bias value close to $b_L=1.7$ at z$\sim$0 we inferred from the halo mass evolution.
However it is likely that more sophisticated models are needed to fully interpret this evolution.
\end{itemize}

Our data provide a robust reference for the clustering of galaxies at an epoch before the peak in star formation rate.
We infer that the star-forming population of galaxies at $z\sim3$ should evolve into the massive and bright ($M_r<-21.5$) galaxy population 
which typically occupies haloes of mass $\langle M_h\rangle = 10^{13.9}$ at redshift $z=0$. More complete modeling seems
necessary to relate high redshift to low redshift galaxy populations,
possibly taking into account non-linear effects, to better explain the galaxy clustering properties at early epochs in
the life of the universe.

\begin{acknowledgements}
We thank Martin Kilbinger for assistance with implementing our HOD model into the CosmoPMC code.
This work is supported by funding from the European Research Council Advanced Grant ERC-2010-AdG-268107-EARLY and by INAF Grants PRIN 2010, PRIN 2012 and PICS 2013. 
AC, OC, MT and VS acknowledge the grant MIUR PRIN 2010--2011.  
This work is based on data products made available at the CESAM data center, Laboratoire d'Astrophysique de Marseille. 
This work partly uses observations obtained with MegaPrime/MegaCam, a joint project of CFHT and CEA/DAPNIA, at the Canada-France-Hawaii Telescope (CFHT) which is operated by the National Research Council (NRC) of Canada, the Institut National des Sciences de l'Univers of the Centre National de la Recherche Scientifique (CNRS) of France, and the University of Hawaii. This work is based in part on data products produced at TERAPIX and the Canadian Astronomy Data Centre as part of the Canada-France-Hawaii Telescope Legacy Survey, a collaborative project of NRC and CNRS.
\end{acknowledgements}


\bibliographystyle{aa}
\bibliography{first_Ania}

\appendix

\begin{figure*}[t]
 \centering
 \includegraphics[width=0.4\textwidth, height=0.9\textwidth, angle =270]{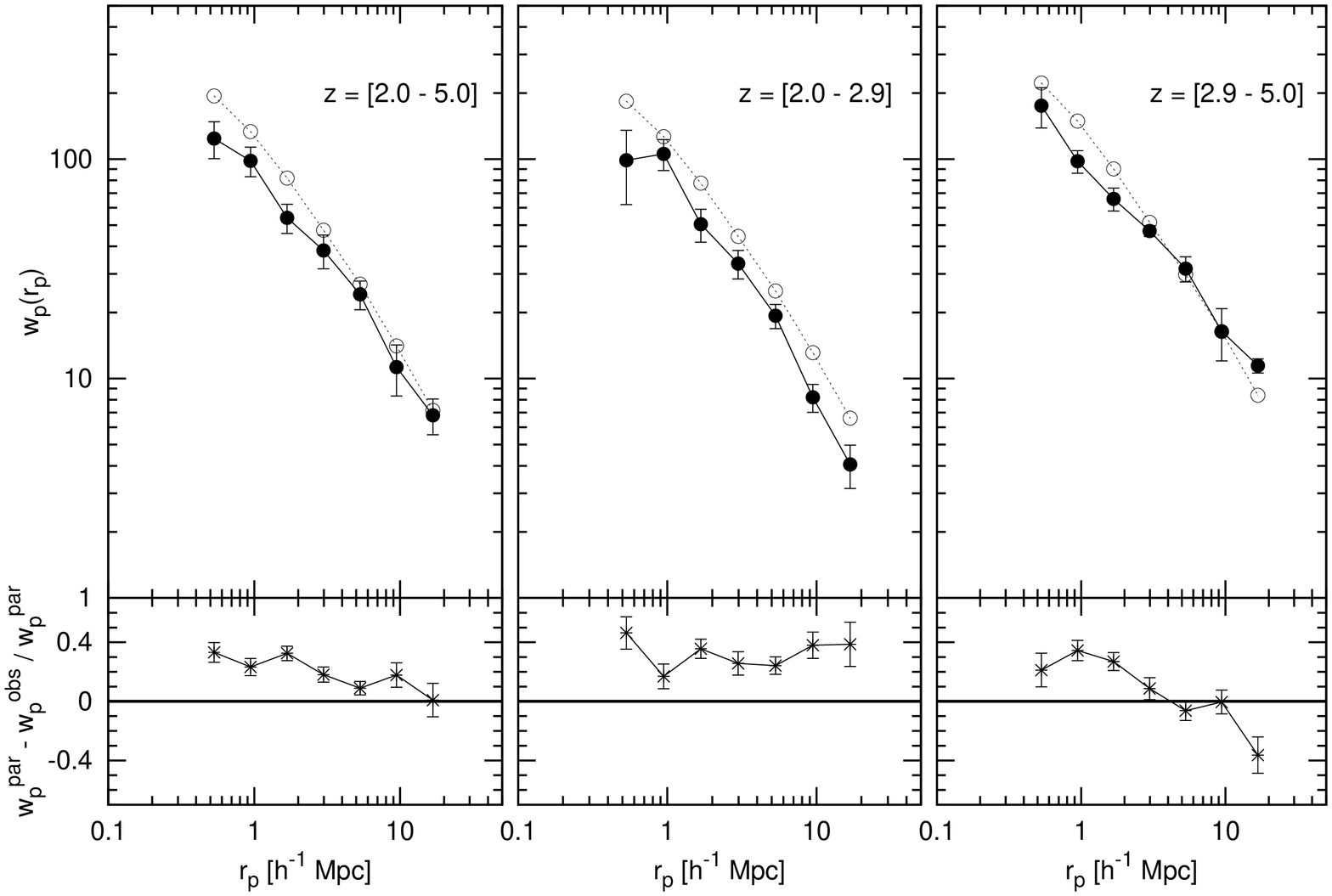}
 \includegraphics[width=0.4\textwidth, height=0.9\textwidth, angle =270]{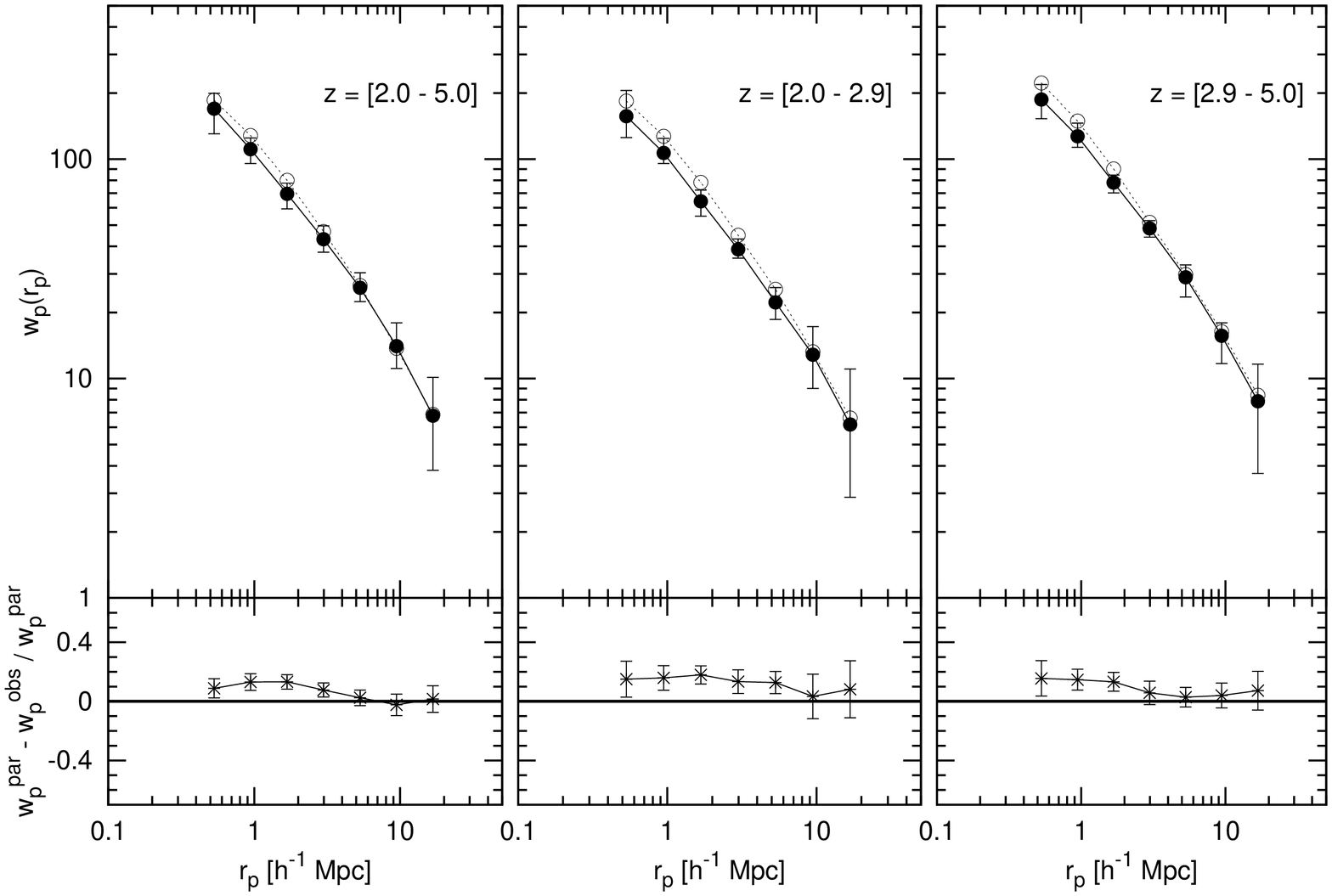}
 \caption{Projected two point correlation function $w_p(r_p)$ measured for 66 VUDS mock catalogs in three redshift ranges. The points correspond to the mean 
measurement of all 66 mock catalogs, while the errors are computed as their standard deviations. 
	  The 'true' $w_p(r_p)$ computed for the whole parent sample (open circles) is compared to that measured from the 'observed' sample (filled circles).
	  \textit{Upper panel:} Without corrections (2) and (3). \textit{Lower panel:} After our full correction scheme was applied.}
 \label{fig:weights_influence}
\end{figure*}

\section{Correlation function corrections scheme}
\label{sec:corr_correction}

In the galaxy surveys, there exists a number of observational biases which may influence the clustering measurements. 
These biases are related to the construction of the survey. 
Both the parent photometric catalogue, from which target galaxies are selected, and the final spectroscopic catalogue may be the source of these biases.
Below, we list the most important survey features that cause the biases. 
Next, we describe the corrections applied to minimize their effects. 
Since the VUDS survey is performed on the same multi-slit spectrograph VIMOS as the VVDS, zCOSMOS and VIPERS surveys, the correction scheme adopted here is based on the same methodology and the ones used for the other VIMOS-based surveys in the past. 
In particular, the main corrections used here have been proposed and fully described by \cite{Pollo2005} and \cite{delaTorre2011}.

The main observational biases in the VUDS data can be listed as follows.

\begin{enumerate}
\item	As shown in Figure \ref{fig:field_galaxy}, some of the areas are excised from the observations, due to the VIMOS lay-out (see \ref{fig:VIMOS_shape}).
	The field of view of this instrument consists of four quadrants separated by $2'$ gaps.  No galaxies are observed in these gaps between quadrants, which influences the pair counts. 
\item	Obviously, not all galaxies from the photometric target candidate sample can be spectroscopically observed. 
	Each galaxy spectrum occupies a certain area on the CCD detector, calculated from the spatial extent of the slit and the length of the
spectrum due to spectral dispersion, which imposes geometrical constraints in target selection as galaxies too close
to each other cannot be targeted simultaneously.  
	It means that the galaxies from the parent photometric sample must be chosen is specific way to effectively allocate as many as possible 'spectral slits' for each observation. 
	In the case of the VIMOS observations, the slit allocation is performed automatically by the Super-SPOC code (\citeauthor{Bottini2005} \citeyear{Bottini2005}).
	As a result, the spectroscopically observed sample is not a random representation of the parent photometric sample, and the introduced bias is relatively complex, especially on the small scales.
\end{enumerate}

These biases, as well as a few other, less important, effects are described in detail by \cite{Pollo2005}. 

However, these biases can be minimized by a combination of corrections, which were presented by \cite{Pollo2005} and \cite{delaTorre2011}, and which we found the best working for the VUDS data:

\begin{enumerate}
\item  Random sample construction.
       The first, most basic part of the correction scheme is the appropriate construction of the random catalog which needs to be geometrically identical (with the exception of the small scale non-linear biasses introduced by the SSPOC) with the spectroscopic sample. 
       Generating random objects in this catalogue, we take into account the shape of the single pointings and quadrants, which are related to the shape of the VIMOS spectrograph.
       Additionally, we exclude the regions removed from the parent photometric sample (e.g. due to the presence of a bright star), by applying the same photometric mask to the random sample.  
       These first-order corrections reduce most of the negative effects on the correlation function.
\item  A global correction. 
       In order to account for the missing pairs (due to the VIMOS limitations and the SSPOC strategy) we assign a "global" weight to each galaxy-galaxy pair.
       Assuming that the parent catalogue is free from angular incompleteness, we define a weighting function $f(\theta)$ as a ratio between the mean number of pairs in the parent photometric catalogue and the main number of pairs in the spectroscopic sample, as a function of angular separation (\citeauthor{delaTorre2011} \citeyear{delaTorre2011}):
	
	\begin{equation}
	  f(\theta) = \frac{1+w_{par}(\theta)}{1+w_{spec}(\theta)},
	\end{equation}
       where $w_{par}{\theta}$ and $w_{spec}(\theta)$ are the angular correlation functions of the parent photometric and spectroscopic samples, respectively. 
       Then, each pair from the spectroscopic sample separated by the angular distance $\theta$ is weighted by this ratio $f(\theta)$. 
\item  Small scale corrections.
       As it comes out, even this strategy does not fully account for the small-scale angular effects of the SSPOC target selection strategy.	
       In order to account for these local small-scale biases we additionally use the local weighting scheme which is also using the parent photometric catalog as a reference. 
       In this case, we count how many galaxies around the targeted galaxy we are missing due to the limited space for spectroscopic slits. 
       Each targeted galaxy pair is then weighted proportionally to its representativeness of the surrounding density field. 
       Similarly to \cite{Pollo2005} we found that the optimal size of the weighting area is the circle with a radius $\sim 40''$.
 \end{enumerate}
 
The correcting scheme including all these ingredients was confirmed to be optimal after a series of tests performed using all 66 VUDS mock catalogues.
In figure \ref{fig:weights_influence} we present the results of the correlation function measurements for the mock catalogues in our three redshift ranges. 
All the points correspond to the means of the results from 66 VUDS mock catalogues, while the errors are computed as the standard deviations between them.
The three upper panels show the comparison between (1) the 'true' (computed for whole mock parent sample) correlation function and (2) the 'observed' correlation function computed only with the most obvious correction number (1), i.e. the appropriate geometrically cut random sample. 
The lower panels present the comparison of the same 'true' correlation function and (3) the 'observed' correlation function when the full correcting scheme described above was applied.

It is clearly seen that the introduced corrections significantly improve the retrieved signal from the 'observed' sample at all the spatial scales, and in all three redshift bins considered.

However, it can be noticed that on small scales ($r_p<3 h^{-1} \textrm{Mpc}$), there is still some signal missing, even after introducing the full optimal correcting scheme (see lower panels of Figure \ref{fig:weights_influence}).
For these separations, our measurement is on average underestimated by $10\%\pm3\%$ with respect to the 'true' value of $w_p(r_p)$ computed from the parent sample.
The possible influence of this effect on the final results is discussed in section \ref{sec:systematics}.
At separations larger than $r_p \sim 3 h^{-1} \textrm{Mpc}$ the value of $w_p(r_p)$ is also slightly underestimated, but only by about $1.6\%\pm0.9\%$. 

Our correction scheme provides a significant improvement in comparison to the average $28\%±5\%$ bias observed without applying it. 
However, the remaining systematic underestimation of the $w_p(r_p)$ has to be taken into account during the correlation function measurements (see Sec. \ref{sec:systematics}).

\begin{figure}
 \centering
 \includegraphics[width=0.4\textwidth]{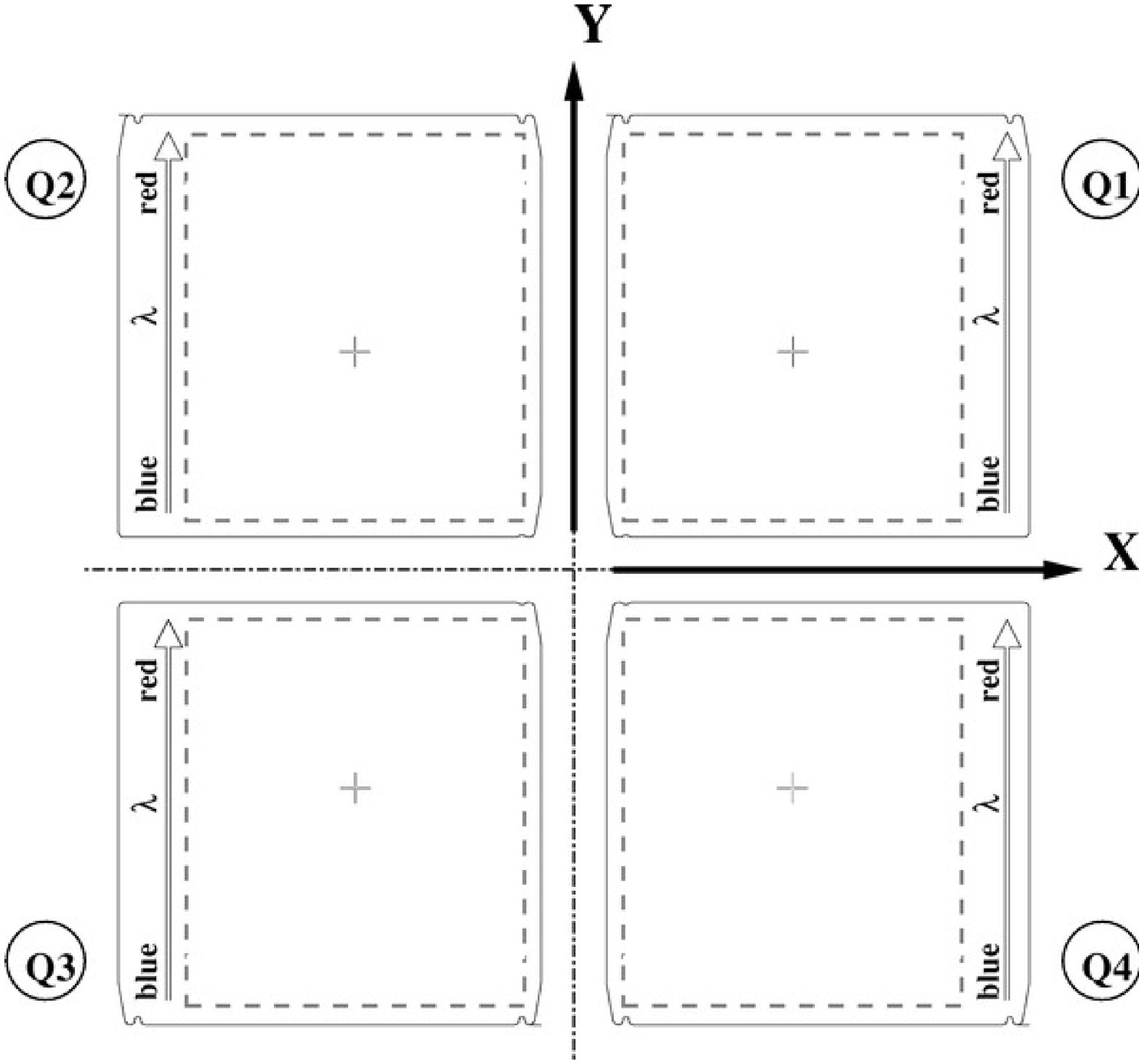}
 \caption{Lay-out of the VIMOS field of view. Picture from \cite{Bottini2005}.}
 \label{fig:VIMOS_shape}
\end{figure}

\section{Projected correlation function measurements}
For every galaxy sub-sample used in this work we present the tables of the projected correlation function measurements $w_p(r_p)$ with associated $1\sigma$ errors at different separations $r_p$ (in units $h^{-1}\textrm{Mpc}$).

\begin{table}
\label{tab:wprp_redshift_all}
 \begin{center}
 \caption{General galaxy population. Projected correlation function measurements of redshift sub-samples.
	  The first column provides the pair-weighted-projected separation of the bin $r_p$. 
	  Subsequent columns provide the projected correlation function values $w_p(r_p)$ along with $1\sigma$ errors.}
 \resizebox{0.5\textwidth}{0.2\textwidth}{
 \begin{tabular}{c|c|c|c} \hline \hline
 \multirow{2}{*}{$r_p$} & \multicolumn{3}{c}{$w_p(r_p)$} \\ \cline{2-4}
			& 	$2.0<z<2.9$		&	$2.0<z<5.0$		&	$2.9<z<5.0$ \\ \hline
    0.29		&	147.40 $\pm$ 42.02	&	107.65 $\pm$ 32.32	&	68.74 $\pm$ 48.22\\
    0.53		&	84.84 $\pm$ 26.39	&	65.03 $\pm$ 21.72	&	67.51 $\pm$ 36.77\\
    0.94		&	46.83 $\pm$ 12.79	&	42.22 $\pm$ 10.26	&	48.77 $\pm$ 15.83\\
    1.68		&	28.71 $\pm$ 7.57	&	28.11 $\pm$ 6.07	&	41.94 $\pm$ 9.80\\
    2.98		&	22.81 $\pm$ 4.87	&	19.20 $\pm$ 4.50	&	20.21 $\pm$ 6.58\\
    5.31		&	12.32 $\pm$ 4.01	&	12.99 $\pm$ 3.06	&	17.87 $\pm$ 4.13\\
    9.44		&	8.63 $\pm$ 2.39		&	8.97 $\pm$ 2.01		&	10.62 $\pm$ 2.82\\
    16.79		&	4.26 $\pm$ 2.08		&	6.23 $\pm$ 1.62		&	9.44 $\pm$ 2.19 \\ \hline \hline
 \end{tabular}}
 \end{center}
\end{table}

\end{document}